\documentclass{emulateapj}

\usepackage{graphicx}

\begin{document}

\title{MIPS 24 Micron Observations of the Hubble Deep Field South: Probing the IR-radio correlation of galaxies at $z > 1$}



\renewcommand{\thefootnote}{\alph{footnote}}

\author{Minh T. Huynh}
\affil{Infrared Processing and Analysis Center, MS220-6, California Institute of Technology,
Pasadena CA 91125, USA. mhuynh@ipac.caltech.edu}

\author{Eric Gawiser}
\affil{Department of Physics and Astronomy, Rutgers University, 136 Frelinghuysen Rd, Piscataway, NJ 08854, USA}

\author{Danilo Marchesini}
\affil{Department of Physics and Astronomy, Tufts University, Medford, MA 02155, USA }

\author{Gabriel Brammer}
\affil{Department of Astronomy, Yale University, New Haven, CT 06520, USAe}

\author{Lucia Guaita}
\affil{Departmento de Astronomia y Astrofisica, Universidad Catolica de Chile, Santiago, Chile}

\begin{abstract}
We present MIPS 24$\mu$m observations of the Hubble Deep Field South taken with the Spitzer Space Telescope. The resulting image is 254 arcmin$^2$ in size and has a sensitivity ranging between $\sim$12 to $\sim$30 $\mu$Jy rms, with a median sensitivity of $\sim$20 $\mu$Jy rms. A total of 495 sources have been cataloged with a signal-to-noise ratio greater than 5$\sigma$. The source catalog is presented as well as source counts which have been corrected for completeness and flux boosting. The IR sources are then combined with MUSYC optical/NIR and ATHDFS radio observations to obtain redshifts and radio flux densities of the sample. We use the IR/radio flux density ratio ($q_{24}$) to explore the IR-radio correlation for this IR sample and find $q_{24}$ = 0.71 $\pm$ 0.31 for sources detected in both IR and radio. The results are extended by stacking IR sources not detected in the radio observations and we derive an average $q_{24}$  for redshift bins between $0 < z < 2.5$. We find the high redshift ($z > 1$) sources have an average $q_{24}$ ratio which is better fit by local LIRG SEDs rather than local ULIRG SEDs, indicating that high redshift ULIRGs differ in their IR/radio properties. So ULIRGs at high redshift have SEDs different from those found locally. Infrared faint radio sources are examined, and while nine radio sources do not have a MIPS detection and are therefore radio-loud AGN, only one radio source has an extreme IRAC 3.6$\mu$m to radio flux density ratio indicating it is a radio-loud AGN at $z > 1$. 

\end{abstract}

\keywords{galaxies: formation --- galaxies: evolution --- galaxies: starburst}

\section{Introduction}

The tight correlation between the far-IR and radio emission of starforming galaxies has been well studied in the local universe and established over many orders of magnitude in luminosity (e.g. \citealp{helou1985,yun2001}). The radio emission of normal galaxies is dominated by synchrotron radiation from relativistic electrons accelerated by supernovae and free-free emission from HII regions \citep{condon1992}. Both supernovae and HII regions are related to young, massive stars ($M \gtrsim10 M_{\odot}$), and hence radio emission in normal galaxies is correlated with star formation. The UV emission from young stars is re-radiated by dust in the mid-to-far-infrared so IR emission is also related to star forming processes. The far-IR/radio correlation is important because it is a probe of the starforming processes taking place. Deviations from the correlation may occur due to, for example, the varying strength of galactic magnetic fields, evolution of dust and metallicity properties, and increasing Inverse Compton losses off the cosmic microwave background (e.g.\citealp{murphy2009, lacki2010}). The far-IR/radio correlation can also be used to select radio loud AGN (e.g. \citealp{donley2005}). Thus, it is important to determine whether the correlation holds at high redshift.

A sensitive infrared probe of high-redshift star formation is the 24 $\mu$m band of the Multiband Imaging Photometer for Spitzer (MIPS) instrument \citep{rieke2004} on board the Spitzer Space Telescope  \citep{werner2004}. Several deep surveys have correlated 24 $\mu$m with radio data to study the mid-IR/radio relation \citep{boyle2007, beswick2008, ibar2008, garn2009}. Some of these authors \citep{boyle2007, beswick2008} observe a change in the mid-IR/radio ratio at faint radio and/or  24 $\mu$m  flux densities, but the differences in the results are likely to be because of selection biases which lead to different populations being studied. 

More recent work in the COSMOS field has found little variation in the IR-radio ratio of starforming galaxies for $z < 1.4$, and for the limited sample of sources confirmed at $z > 2.5$ the average IR-radio ratio is still the same as that found locally \citep{sargent2010}. Discrepant measurements of the average IR-radio ratio can be reconciled by the fact that previous studies selected in either IR or radio, or required a detection in both bands \citep{sargent2010}. 

In 1998 the {\sl Hubble Space Telescope (HST)} observed a region in the southern continuous viewing zone as a complementary observation to the Hubble Deep Field North (HDF-N, \citealp{williams1996}). This field, known as the Hubble Deep Field South (HDF-S; \citealp{williams2000}), reaches a limit of $\sim$30 magnitude in the four {\sl HST}  WFPC2 broadband filters. Simultaneous observations with the other HST instruments, NICMOS and STIS, reach similarly deep levels in the near-infrared and ultraviolet. In addition to the main {\sl HST} deep fields, a mosaic of nine flanking fields were imaged to shallower levels of $\sim$25 magnitude in I (F814W) \citep{lucas2003}. Wide-field ground based imaging has been performed by \cite{palunas2000} and the Multiwavelength Survey by Yale-Chile (MUSYC) team \citep{gawiser2006}, reaching R band magnitude limits of $\sim$25 and 26, respectively. This field has been the target of deep radio observations reaching $\sim$10$\mu$Jy rms at 1.4, 2.5, 5.2 and 8.7 GHz \citep{huynh2005, huynh2007}. 

Confirming the COSMOS results requires a deep and reasonably wide MIPS field. In this paper we present the MIPS 24$\mu$m observations of the Hubble Deep Field South and take advantage of the enormous amount of multiwavelength data available to study the IR-radio correlation of galaxies detected in this field. We use a stacking technique in the radio to determine the average IR-radio ratio of a flux-limited IR selected galaxy sample.

In Section 1 we describe the MIPS observations and data reduction. We summarize the multiwavelength data used for this work, in particular the MUSYC data, in Section 2. The MIPS source detection and extraction is discussed in Section 4 and here we present the 24 $\mu$m catalog, source counts and redshift distributions. Section 5 reports our work on the IR-radio correlation in the HDFS  and our identification of infrared-faint radio sources in the field. We conclude with a summary in Section 6.  A Hubble constant of $71\,{\rm km}\,{\rm s}^{-1}{\rm Mpc}^{-1}$, and a standard $\Lambda$-CDM cosmology with $\Omega_{\rm M}=0.27$ and $\Omega_{\rm \Lambda}=0.73$ is assumed throughout this paper.

\section{MIPS Observations and Data Reduction}

The MIPS 24 $\mu$m observations of the Hubble Deep Field South were carried out over 2 campaigns. The first was in 2004 as part of the Guaranteed Time Observations (PI Fazio) and the second in 2007 Cycle 3 observations (PID30873, PI Labbe ). The GTO observations were taken using the MIPS Photometry mode and covered six pointings in the HDF-S region. The Cycle 3 observations consisted of four pointings which covered the area west of the WFPC2 field (Figure 1). The coverage (Figure 1) is very uneven, with effective on sky integration time per pixel ranging from about 4500 seconds in the deepest area to 900 seconds in the shallow regions. 

The Basic Calibrated Data (BCDs) were downloaded from the {\sl Spitzer} Science Center (SSC) archive and were reduced using version 18.1.5 of the MOsaicker and Point source EXtractor (MOPEX) software from the SSC. As recommended, the first frame from each commanded sequence of exposures (i.e. DCENUM=0) was discarded.  Mosaicing the BCDs at this point resulted in latent artifacts. Flatfields were created using the MOPEX routine {\em flatfield.pl}, which makes a median flatfield from a sequence of BCDs. For the GTO data we found that bright sources in the field affected the flatfield generation because of insufficient dithering within the individual Astronomical Observing Requests (AORs) in the GTO data. Therefore we combined all AORs in the GTO data to reduce the effect of bright sources being counted in the flatfield statistics. All four GTO AORs in 2004 were observed within four hours of each other, so combining all them to make a master flat is reasonable because the instrument flatfield characteristics do not change over such a short timescale.  The Cycle 3 observations do not have the dithering issue, and flatfields were generated for each AOR from simply feeding the BCDs of the individual AOR into {\em flatfield.pl}. 

After flatfielding the BCDs the  MOPEX task {\em overlap.pl} was used to level the backgrounds. The BCDs were then mosaicked in the usual way with MOPEX  using the default pixel scale of 2.45" per pixel. The outlier rejection, which has a default of 5$\sigma$, was set to the more aggressive value of 3$\sigma$. The resulting image, with units of MJy/sr, is shown in Figure 1. This image is about 254 arcmin$^2$ in size. 

\section{Multiwavelength ancillary data}

The field was observed by the main three instruments on the Hubble Space Telescope in 1998 to depths rivaling that of original Hubble Deep Field in the north \citep{williams2000}.  The {\sl HST} imaging reaches 10$\sigma$ AB magnitudes of approximately 27.7 in I band (F814W) \citep{casertano2000} in the deepest region. Including the flanking fields, Hubble imaging covers only a small $\sim$50 arcmin$^2$ area, however. 

More recently, the Multiwavelength Survey by Yale-Chile (MUSYC) observed an approximate 37 $\times$ 37 arcmin region surrounding the Hubble Deep Field South in UBVRIz' optical bands \citep{gawiser2006}. These observations reach UBVRIz' magnitude limits (5$\sigma$, AB) of 26.0, 26.1, 26.0, 25.8, 24.7 and 23.6, respectively. The MUSYC team also obtained NIR imaging in J, H and K' filters reaching 5$\sigma$ depths of 22.9, 21.8 and 21.1 Vega magnitudes over two 10 $\times$ 10 arcmin fields in the extended Hubble Deep Field South \citep{quadri2007}. 

Infrared Array Camera (IRAC) data from {\sl Spitzer} was also obtained as part of the GTO (PI Fazio) and 2007 Cycle 3 observations (PID30873, PI Labbe ). The BCDs from the SSC archive had the usual procedures applied to remove cosmics rays and artifacts, such as column pulldown, muxbleed and maxstripe (see \citealp{marchesini2009} for details). Individual background-subtracted images were combined into mosaics covering two 10 $\times$ 10 arcmin regions to limiting AB magnitudes (3$\sigma$) of 24.5, 24.0, 22.2, and 22.1 at 3.6, 4.5, 5.8 and 8.0 $\mu$m, respectively\citep{marchesini2009}.

Accurate photometric redshifts were determined with \textsc{eazy} from the MUSYC 14 band optical-IRAC data \citep{marchesini2009}. \textsc{eazy}, a template-fitting code similar to HYPERZ \citep{bolzonella2000} and BPZ \citep{benitez2000}, has been found, has been found to provide high quality redshifts over $0 \lesssim z \lesssim 4$ \citep{brammer2008}, despite the fact that it is not trained on a spectroscopic sample.  \textsc{eazy} has the option of linearly combining templates when fitting sources and provides multiple estimates of the photometric redshift, including $z_{peak}$, which is determined from the full probability distribution rather than a straight $\chi^2$ minimization. In the case where there are two widely-separated peaks in the redshift probability function, $z_{peak}$ selects the peak with the largest integrated probability. In the eCDFS the MUSYC team achieved photometric redshift accuracies of $\delta z / (1 + z)$ $\sim$ 0.008 with \textsc{eazy}, but this result came from supplementing UBVRIz'JHK optical imaging, with {\sl Spitzer} IRAC and 18 medium band optical images from Subaru (Cardamone et al. submitted). The photometric redshifts from the 14 band optical-IRAC data have an accuracy of $\delta z / (1 + z) = 0.000 \pm 0.033$ and the fraction of catastrophic outliers is only 4\% \citep{marchesini2009}.

The radio observations of the HDF-S were made between 1998 and 2001 with the Australia Telescope Compact Array (ATCA) using all four available frequency bands. Between 100 and 300 hours of observing at each band yielded images at 1.4, 2.5, 5.2, and 8.7 GHz with maximum sensitivities of $\sim$10 $\mu$Jy rms \citep{huynh2005,huynh2007}. 

\section{MIPS Image Analysis}

\subsection{Source Detection and Extraction}

Sources detection and extraction was performed using the Astronomical Point Source extraction (APEX) tools in the MOPEX software. First, APEX performs a local background subtraction by taking a median within a box surrounding each pixel. The box size was set to 34 pixels, or $\sim$14 times the Full Width Half Maximum (FWHM) of the 24$\mu$m PSF, and the median calculated after rejecting the 100 highest  pixels. APEX detects sources when a pixel cluster has values which are above the noise threshold, but this requires a good estimate of the noise. The MOPEX and APEX software can produce several types of noise images. The ``std" noise image produced at the mosaicking stage provides the empirical scatter of each pixel from the repeated observations divided by the square-root of the number of observations. This underestimates the true noise because it doesn't account for pixel-to-pixel correlated noise or confusion noise. APEX generates an output ``noise" map by calculating the noise within a local box, and here we use the same settings as for the background subtraction - 34 pixel rectangular box and rejection of 100 outlier pixels.  Using an approach similar to \cite{frayer2009}, the ``std" and ``noise" images are then combined in quadrature with equal weights to obtain a combined uncertainty image for source detection and extraction. 

We then checked whether the combined uncertainty image was reasonable by checking this against an independent total noise value. An independent estimate of the total noise in the image was made by measuring the flux density in apertures randomly placed on the image and calculating the standard deviation of the measurements. Two hundred apertures were placed randomly in 60 by 60 pixels regions of a residual image with bright sources removed. The standard deviation of the flux density measurements agreed with the combined uncertainty image to within 50\%.

Sources were detected by APEX if they had a peak with a signal-to-noise ratio (SNR) greater than 4 and were fitted using the Point-source Response Function (PRF) image provided by the SSC. The sources were then included in the final catalog if they had a fitted SNR greater than 5. This relatively high SNR was chosen to minimize the number of spurious detections. Even with this SNR ratio the initial source list (504 total) was visually examined for potential spurious sources such as those in the Airy rings of bright sources. Only 9 ($\sim$2\%) sources were removed, suggesting that the APEX algorithm is reliable. The final 24 $\mu$m catalog contains 495 sources. 

The full $5\sigma$ catalog is presented in Table \ref{catalog}. A description of the Table is as follows. 

{\em Column (1)} --- Source ID.

{\em Column (2)} --- Right Ascension in J2000.
 
{\em Column (3)} --- Declination in J2000.

{\em Column (4)} --- 24 $\mu$m flux density, in $\mu$Jy.

{\em Column (5)} --- 24 $\mu$m flux density uncertainty, in $\mu$Jy. This is the formal error from APEX.

{\em Column (6)} --- Signal to noise ratio of the source.

\subsection{Completeness and Flux Boosting}

The completeness of the source catalog was estimated by simulations. Monte-Carlo-like simulations were performed by injecting 10000 sources at random locations on to the map and extracting them using the same technique as adopted for the production of the catalog. While completeness levels vary significantly across the image as a function of coverage and flux density, we can recover the overall completeness level of the generated catalog by injecting sources over the full image. A single source was injected per simulation, and the input flux density varied from 50 to 3000 $\mu$Jy to sample the full range of interest. The completeness as a function of flux density is shown in Figure 2.  The completeness rises steeply from about 10\% at 100 $\mu$Jy to approximately 90\% at 250 $\mu$Jy. The 50\% completeness level occurs at approximately 150 $\mu$Jy. 

Source detection algorithms that rely on finding a peak above a local noise background lead to flux boosting. This is because sources that lie on a noise peak have their flux density increased and therefore have a higher probability of being detected, while sources which lie on a noise trough have decreased flux densities and may be excluded altogether. This effect is most pronounced in the faintest flux density bins, i.e. in sources with a SNR close to the limit of the catalog. The degree of flux boosting can be estimated by examining the output to input flux density of the simulations (Figure 2). We find that flux densities are boosted by about 1\% to 4\% for measured flux densities of 100 to 300 $\mu$Jy, on average. The flux boosting is negligible for sources with true flux densities brighter than 300 $\mu$Jy.

The simulations can also be used to gauge the positional accuracy of the catalog by comparing input and output positions. The median of the RA and Dec offsets are plotted for various input flux density bins in Figure 3. The positional accuracy can be estimated from the standard deviation in the offsets of each of these bins. We find that at the faintest levels (80 $\mu$Jy) the RA and Dec uncertainties are approximately 0.8 and 1.3 arcsec, respectively. The positional accuracy is much better than 0.5 arcsec for sources that are 1 mJy and brighter. This is the random component only, and doesn't take into account any systematic offsets in the pointing. {\sl Spitzer} has an absolute pointing uncertainty of about 1 arcsec. 

\section{Results}

\subsection{Source Counts}

The source counts were derived from the SNR $\geq$ 5 catalog. We correct the source counts for flux boosting and completeness using the results of Section 4.2.  Figure 4 shows the Euclidean normalized source counts ($dN/dS \times S^{2.5}$) and the results are tabulated in Table 2.  The uncertainties listed are derived from Poissonian statistics. The counts from this work span the peak near 0.2 -- 0.3 mJy. We have also plotted the counts from five Guaranteed Time Observation (GTO)  fields \citep{papovich2004}, the FIrst Look Survey (FLS) field \cite{marleau2004} and Spitzer Wide-field InfraRed Extragalactic (SWIRE) fields \cite{shupe2008} for comparison. In the fainter bins HDFS counts are broadly consistent with that from previous work, showing the steep decline for $S_{24}$ $<$ 0.2 mJy. At brighter flux densities the counts are consistent with the FLS, within the uncertainties, but lower than that from the GTO or SWIRE fields. At $\sim$0.5 mJy the SWIRE counts are about 40\% higher the HDFS, and about 20\% higher than the FLS or GTO source counts. This illustrates that there is significant field to field variation, which \cite{shupe2008} estimated to be about 10\% for sub-mJy flux densities in regions as large as the SWIRE fields ($>$ 7 degrees). 

Also overlaid in Figure 4 are the normalized counts from the model of \cite{lagache2004}. See \cite{lagache2004} for details, but briefly, this model uses starburst and normal galaxy templates and fits for the evolution of the local luminosity function that is best constrained by observed source counts and redshift distributions. We note that this model used counts at 15, 60, 170 and 850 $\mu$m from literature as well as the 24 $\mu$m counts from \cite{papovich2004}. The model is broadly consistent with the HDFS, FLS and GTO counts and reproduces the turnover at $\sim$0.1 mJy that is seen in these fields.

\subsection{Optical Counterparts}

We used the MUSYC catalogs to identify optical counterparts to the MIPS sources and obtain redshift information from either the spectroscopic or photometric catalogs. The MUSYC imaging overlaps almost the entire MIPS imaging, with only the $\sim$ 24 arcmin$^2$ region north of Dec = $-$60.4994 falling outside the MUSYC observations. Thus, 25/495 (5\%) MIPS sources do not have MUSYC coverage. The MUSYC K-selected catalog (Quadri et al. 2007) and v1.0 optical catalog \citep{gawiser2006} were searched for counterparts to the rest of the MIPS sample. 

With the $\sim$1 arcsec seeing of the observations and the astrometric uncertainties of the USNO-A2 stars used for astrometry, the positional uncertainties of the MUSYC sources are about 0.2 -- 0.3 arcsec. For MIPS the 1$\sigma$ positional uncertainty can be estimated by $0.5 \theta_{\rm FWHM}$/ (S/N) assuming Gaussian noise, and hence a beam size of 6 arcsec implies a positional uncertainty of 0.6 arcsec (in each coordinate) for the faintest sources. Including the {\sl Spitzer} 1 arcsec absolute pointing uncertainty, the 1$\sigma$ offset between MIPS and MUSYC sources should be about 1.3 arcsec for the faintest MIPS sources. 

Figure 5 shows a plot of distribution of the offsets between all MIPS and MUSYC K-selected (Quadri et al. 2007) sources within 2 arcsec. As expected, there is a peak in the distribution of offsets at $\lesssim$0.5 arcsec. The contribution from chance matches due to the optical source density is also shown in Figure 5. We find that the number of chance matches meets the number of observed matches at a radius of $\sim$ 1.5 arcsec and therefore use this as our matching radius. 369 MIPS sources have a counterpart in the K-selected catalogs and a photometric redshift from \cite{marchesini2009}. A further 37 have counterparts in the optical MUSYC catalogs, which cover a larger area than the K band imaging. For these 37 sources \textsc{eazy} photometric redshifts were calculated from the MUSYC UBVRIz optical data plus a narrow band 5000 ûA image. A comparison of these seven-band photometric redshifts with 166 spectroscopic redshifts from literature finds only 18\% have $\delta z / (1 + z) >  0.2$. Excluding these outliers, we find $\delta z / (1 + z) = 0.027 \pm 0.046$, and thus conclude the photometric redshifts are statistically reliable.  In total 406/495 (82\%) of the MIPS sources have a MUSYC optical or NIR counterpart. Three have more than one MUSYC source within 1.5 and in these cases the optical candidates were individually examined and the most likely candidate chosen using the P statistic (e.g. \citealp{downes1986, lilly1999}). 

Spectroscopic redshifts from the literature are available for 22 of the MIPS sources, and we assume the photometric redshift, where available, for the remaining sources. The redshift distribution of the MIPS sources peak at $z \sim 0.5$ with a tail to high redshift ($z > 2$, Figure 5). The total infrared luminosity (8 -- 1000 $\mu$m) of the sources is estimated by fitting the 24 $\mu$m flux density to the luminosity dependent SED templates of \cite{ce01}. We integrate the best fit template over 8 -- 1000 $\mu$m to derive the total IR luminosity (LIR) (Figure 6). We find luminous IR galaxies (LIRGS, $10^{11} L_\odot< $ LIR $< 10^{12} L_\odot$) are detected out to $z \sim 1.1$, while ultraluminous IR galaxies (ULIRGs, LIR $> 10^{12} L_\odot$) are detected out to higher redshifts. We note that while the \cite{ce01} templates exhibit observed FIR flux density ratios which are more representative of  $z \sim 1$ galaxies than Dale and Helou (2002) or Lagache et al. (2003) templates (Magnelli et al. 2009), the choice of templates can lead to $\sim0.3$ dex variation in luminosity.  Fitting to the 24 $\mu$m flux density alone can give a good LIR estimate up to about $z \sim 1.4$, but for $1.4 < z < 2.5$ the LIR can be overestimated by up to a factor of four (Murphy et al. 2009).

\section{Discussion}

\subsection{Infrared-Radio Correlation}

Radio counterparts to the MIR sources were obtained using the 1.4 GHz radio catalog of the HDFS \citep{huynh2005}. We applied a simple radial match of 3 arcsec, and find 84 MIPS sources have a radio counterpart. The matching distance is large enough for obtaining bona-fide matches since the 24 $\mu$m positional uncertainties are $\sim 1.2$ arcsec and the radio positions are better than about 1 arcsec in most cases (Huynh et al. 2005). The P-statistic suggests the probability of one or more radio sources within 3 arcsec of a IR one by chance is $\sim$ 0.3\%, but this estimate does not take into account the individual positional uncertainties and assumes the radio source population is not clustered. 

The IR/radio correlation as a function of redshift is presented in Figure 7, shown as the observable parameter $q_{24}$ = $\log (S_{\rm 24\mu m}) / S_{\rm 1.4 GHz})$. For the 84 sources with detections in both the IR and radio the median $q_{24}$ = 0.71 $\pm$ 0.31. This is consistent with the $q_{24}$ = 0.84 $\pm$ 0.28 found in the xFLS \cite{appleton2004} and $q_{24}$ = 0.66 $\pm$ 0.39 in the SubaruÐXMMÐNewton Deep Field (SXDF, \citealp{ibar2008}). Overlaid on Figure 7 are SED tracks from local templates by \cite{ce01}. We find that up to redshift one the IR sources appear to follow tracks of local SEDs, suggesting the IR-radio correlation holds to this redshift. Above redshift one the 24 $\mu$m band starts to observe the MIR regime ($\lesssim 12\mu$m restframe) where PAH emission and silicate absorption strengths vary widely in individual galaxies, and hence a wide range of MIR/radio ratios is possible. The high redshift ($z > 1$) sources in Figure 7 also fall within the ratios allowed by local templates, but most are best described by the SEDs of relatively low luminosity SEDs (L $< 10^{12}$ Lsun). These sources have luminosities that are ULIRG-like ($>10^{12}$ Lsun, Figure 6) but exhibit ratios similar to local $\sim 10^{11}$ Lsun LIRGs, implying that the SEDs of high redshift ULIRGs differ to that of local ones.  We speculate that these ULIRGs have strong PAH emission, as seen in other IR samples of high redshift ULIRGs (e.g. \citealp{yan2007}), but IR spectra are needed to confirm that hypothesis.

To extend the results we stack the IR sample in the radio data. First, the IR sources were binned into 6 redshifts bins, as shown in Table 3.  For each redshift bin, the radio data was stacked at positions of all undetected sources and a noise-weighted mean image was produced. Stacks of width 60\arcsec\ wide were produced and the photometry performed by fitting a point source to the stacked radio image. All stacks, shown as postage stamps in Figure 8, have a significant detection. The resulting $q_{24}$ values from the stacked radio flux densities are shown as blue squares in Figure 7. They have been combined with the actual detections to derive an average $q _{24}$ for all MIPS sources in the various redshift bins (green squares). We find that the average observed  $q_{24}$ at high redshift ($z > 1$) is approximately 0.86 (Table 3) and this is consistent with local LIR $\sim 10^{11}$ LIRG SED (Figure 7). At these redshifts our sample have ULIRG luminosities or greater, so this is further evidence that the average IR luminous galaxy at high redshift is better described by a lower luminosity local template. This result is consistent with complementary studies which have found that some high redshift  IR luminous galaxies have cooler SEDs than local galaxies of similar luminosities \citep{seymour2010, muzzin2010}.

As mentioned before, the high redshift ULIRGs may exhibit stronger PAH emission than local galaxies, leading to a greater $q_{24}$ then expected from local SEDs.
However, the increased $q_{24}$ at high redshift may also result from suppressed radio emission. If a galaxy has just begun a burst of star formation then the MIR dust emission may be enhanced due to star formation but the supernova rate, which powers the radio emission, lags the star formation by about $\sim 30$ Myr (the approximate lifetime of 8 M$_\odot$ stars). It is possible that some of the sources at high redshift are nascent starbursts. Increased inverse Compton scattering of the cosmic microwave background at high redshift can lead to lower radio emission than in the local universe, and since the energy density of the CMB scales as $(1 + z)^4$ this could be significant at high redshifts. Another potential explanation is that the high redshift galaxies have lower magnetic field strengths than local galaxies, leading to decreased electron confinement and therefore lower radio emission. 

\subsection{Infrared-Faint Radio Sources}

Infrared-Faint Radio Sources (IFRSs) are a class of radio sources recently discovered in the Australia Telescope Large Area Survey (ATLAS)
by \cite{norris2006}. These are radio sources brighter than a few hundred $\mu$Jy at 1.4 GHz which have no observable infrared counterpart in the Spitzer Wide-area Infrared Extragalactic Survey (SWIRE; \citealp{lonsdale2004}).  Most have flux densities of a few hundred $\mu$Jy at 1.4 GHz, but some are as bright as 20 mJy, resulting in extreme infrared to radio ratios. Recent VLBI detections have constrained the radio sources sizes to less than 0.03 arcsec, suggesting IFRSs are compact AGN \citep{norris2007,middelberg2008}. Deeper Spitzer legacy survey data in the extended Chandra Deep Field South yielded two IRAC detections of the four IFRSs in that field, and SED modeling of these sources with the new constraints showed that they are consistent with $z > 1$ radio-loud AGN \citep{huynh2010}. 

The MIPS data of the Hubble Deep Field South allows us to search for IFRSs in this field. There are nine ATHDFS sources with $S_{1.4GHz} > 200$ $\mu$Jy that do not have a MIPS counterpart. These are listed in in Table 4 along with their MIPS ($5 \sigma$) limit and IRAC properties. All of these sources have low $q_{24}$ values consistent with radio-loud AGN, and most have moderate photometric redshifts of $0.417 < z < 1.548$. Most are also relatively bright in the IRAC channels, and would have been detected by SWIRE depth IRAC imaging. Of the nine sources, one (ATHDFS\_J223205.9$-$603857) is not in the $5 \sigma$ IRAC catalog of \cite{marchesini2009} and one is not in the IRAC field. Hence only one source matches the Norris et al. 2006 criteria for IFRS.  One IFRS in the 252 arcmin$^2$ MIPS HDFS region is consistent with the IFRS source density of $\sim$ 16/deg$^2$ in the eCDFS.  ATHDFS\_J223205.9$-$603857 has a $S_{3.6{\mu}m}$ (5$\sigma$) limit of 2 $\mu$Jy and hence a radio-to-IRAC flux density ratio ($S_{1.4GHz}$/$S_{3.6{\mu}m}$) greater than about 130.  Similarly high radio-to-IRAC flux density ratios are found for both IFRS and high redshift radio galaxies (HzRGs) \citep{seymour2007, huynh2010}, suggesting they are high-redshift AGN. For ATHDFS\_J223205.9$-$603857, the low value of $q24$ ($< -0.50$) and high value of $S_{1.4GHz}$/$S_{3.6{\mu}m}$ therefore suggests it is a high redshift ($z > 1$) radio-loud AGN. 

The radio spectral index between 1.4 and 2.5 GHz for these nine ATHDFS sources is also listed in Table 4. All but two have radio SEDs with a spectral index $\alpha \sim -0.8$ ($S \propto \nu^\alpha$), which is consistent with synchrotron emission. The IFRS candidate ATHDFS\_J223205.9$-$603857 has a spectral index $\alpha = -0.80 \pm 0.29$. This result is consistent with Middelberg et al. 2010 (submitted) who find IFRSs have $\alpha < -0.7$.

\section{Summary}

We have presented the MIPS 24 $\mu$m observations of the Hubble Deep Field South from the {\sl Spitzer} Space Telescope. This new MIR image, 250 arcmin$^2$ in area,  reaches a maximum sensitivity of $\sim$12 $\mu$Jy rms, with the bulk of the image achieving $\sim$ 20 $\mu$Jy rms. A catalog of sources brighter than 5$\sigma$ was compiled, comprising 495 sources. The completeness corrected and flux-deboosted 24 $\mu$m source count was presented, and while the counts are consistent with results in the literature they show that there is significant field to field variation. 

The MIPS sources were matched to MUSYC photometric redshift and ATHDFS 1.4 GHz radio catalogs. The IR-radio ratio $q_{24}$,  $\log (S_{\rm 24\mu m}) / S_{\rm 1.4 GHz})$, for 84 sources with both IR and radio detections have a median  $q_{24}$ = 0.71 $\pm$ 0.31, consistent with previous results. We find that while there is diversity in galaxy SEDs, the IR-radio correlation appears to hold out to $z \sim 1$. The results were extended by splitting the sample into redshift bins and stacking all IR sources not detected in the radio to derive a an average radio flux density and hence an average $q_{24}$. The observed average $q_{24}$ was compared with that expected from local SED templates and at $z > 1$ the observed average $q_{24}$ is found to be better described by LIRG templates, even though these sources have ULIRG-like luminosities. This is further evidence that the average IR luminous galaxy at high redshift is better described by a lower luminosity local template. The change in $q_{24}$  may be due to enhanced MIR emission, possibly from hot dust heated by AGN or relatively bright PAH emission, or suppressed radio emission due to lower strength magnetic fields, inverse Compton scattering off the CMB. Several of the high redshift galaxies may be nascent starburst where the radio emission from supernovae is delayed compared to the IR and hence a high  $q_{24}$ results. More detailed IR observations and spectroscopy to determine the AGN/SB nature of the IR galaxies is required to differentiate between these possible scenarios. 

The {\sl Herschel Space Observatory}, launched in May 2009, is currently undertaking deep IR surveys in bands ranging from 70 to 450 $\mu$m. These bands are close to the peak of the IR emission from dust and not significantly affected by AGN or PAH emission. With its unprecedented depth and resolution, {\sl Herschel} will observe large complete samples of IR galaxies to redshifts of $z = 2$ and beyond. In combination with upcoming radio facilities such as eVLA, ASKAP, and MeerKAT, this will allow a major improvement in the study of the IR-radio correlation in the high redshift universe. 
 
\acknowledgements 

MTH wishes to thank D. Frayer for useful discussions. This work is based in part on observations made with the {\sl Spitzer Space Telescope}, which is operated by the Jet Propulsion Laboratory, California Institute of Technology under a contract with NASA.  Support for this work was provided by NASA through an award issued by JPL/Caltech. This material is also based on work supported by the National Science Foundation under grant AST-0807570.

\bibliographystyle{aj}
\bibliography{paper_refs}

\begin{thebibliography}{}

\bibitem[\protect\citeauthoryear{{Appleton} et~al.}{{Appleton}
  et~al.}{2004}]{appleton2004}
{Appleton}, P.~N., et~al. 2004, \apjs, 154, 147

\bibitem[\protect\citeauthoryear{{Ben{\'{\i}}tez}}{{Ben{\'{\i}}tez}}{2000}]{be%
nitez2000}
{Ben{\'{\i}}tez}, N. 2000, \apj, 536, 571

\bibitem[\protect\citeauthoryear{{Beswick} et~al.}{{Beswick}
  et~al.}{2008}]{beswick2008}
{Beswick}, R.~J., {Muxlow}, T.~W.~B., {Thrall}, H., {Richards}, A.~M.~S.,  \&
  {Garrington}, S.~T. 2008, \mnras, 385, 1143

\bibitem[\protect\citeauthoryear{{Bolzonella}, {Miralles}, \&
  {Pell{\'o}}}{{Bolzonella} et~al.}{2000}]{bolzonella2000}
{Bolzonella}, M., {Miralles}, J.,  \& {Pell{\'o}}, R. 2000, \aap, 363, 476

\bibitem[\protect\citeauthoryear{{Boyle} et~al.}{{Boyle}
  et~al.}{2007}]{boyle2007}
{Boyle}, B.~J., {Cornwell}, T.~J., {Middelberg}, E., {Norris}, R.~P.,
  {Appleton}, P.~N.,  \& {Smail}, I. 2007, \mnras, 376, 1182

\bibitem[\protect\citeauthoryear{{Brammer}, {van Dokkum}, \& {Coppi}}{{Brammer}
  et~al.}{2008}]{brammer2008}
{Brammer}, G.~B., {van Dokkum}, P.~G.,  \& {Coppi}, P. 2008, \apj, 686, 1503

\bibitem[\protect\citeauthoryear{{Casertano} et~al.}{{Casertano}
  et~al.}{2000}]{casertano2000}
{Casertano}, S., et~al. 2000, \aj, 120, 2747

\bibitem[\protect\citeauthoryear{{Chary} \& {Elbaz}}{{Chary} \&
  {Elbaz}}{2001}]{ce01}
{Chary}, R.,  \& {Elbaz}, D. 2001, \apj, 556, 562

\bibitem[\protect\citeauthoryear{{Condon}}{{Condon}}{1992}]{condon1992}
{Condon}, J.~J. 1992, \araa, 30, 575

\bibitem[\protect\citeauthoryear{{Donley} et~al.}{{Donley}
  et~al.}{2005}]{donley2005}
{Donley}, J.~L., {Rieke}, G.~H., {Rigby}, J.~R.,  \& {P{\'e}rez-Gonz{\'a}lez},
  P.~G. 2005, \apj, 634, 169

\bibitem[\protect\citeauthoryear{{Downes} et~al.}{{Downes}
  et~al.}{1986}]{downes1986}
{Downes}, A.~J.~B., {Peacock}, J.~A., {Savage}, A.,  \& {Carrie}, D.~R. 1986,
  \mnras, 218, 31

\bibitem[\protect\citeauthoryear{{Frayer} et~al.}{{Frayer}
  et~al.}{2009}]{frayer2009}
{Frayer}, D.~T., et~al. 2009, \aj, 138, 1261

\bibitem[\protect\citeauthoryear{{Garn} \& {Alexander}}{{Garn} \&
  {Alexander}}{2009}]{garn2009}
{Garn}, T.,  \& {Alexander}, P. 2009, \mnras, 394, 105

\bibitem[\protect\citeauthoryear{{Gawiser} et~al.}{{Gawiser}
  et~al.}{2006}]{gawiser2006}
{Gawiser}, E., et~al. 2006, \apjs, 162, 1

\bibitem[\protect\citeauthoryear{{Helou}, {Soifer}, \&
  {Rowan-Robinson}}{{Helou} et~al.}{1985}]{helou1985}
{Helou}, G., {Soifer}, B.~T.,  \& {Rowan-Robinson}, M. 1985, \apjl, 298, L7

\bibitem[\protect\citeauthoryear{{Huynh}, {Jackson}, \& {Norris}}{{Huynh}
  et~al.}{2007}]{huynh2007}
{Huynh}, M.~T., {Jackson}, C.~A.,  \& {Norris}, R.~P. 2007, \aj, 133, 1331

\bibitem[\protect\citeauthoryear{{Huynh} et~al.}{{Huynh}
  et~al.}{2005}]{huynh2005}
{Huynh}, M.~T., {Jackson}, C.~A., {Norris}, R.~P.,  \& {Prandoni}, I. 2005,
  \aj, 130, 1373

\bibitem[\protect\citeauthoryear{{Huynh} et~al.}{{Huynh}
  et~al.}{2010}]{huynh2010}
{Huynh}, M.~T., {Norris}, R.~P., {Siana}, B.,  \& {Middelberg}, E. 2010, \apj,
  710, 698

\bibitem[\protect\citeauthoryear{{Ibar} et~al.}{{Ibar} et~al.}{2008}]{ibar2008}
{Ibar}, E., et~al. 2008, \mnras, 386, 953

\bibitem[\protect\citeauthoryear{{Lacki} \& {Thompson}}{{Lacki} \&
  {Thompson}}{2010}]{lacki2010}
{Lacki}, B.~C.,  \& {Thompson}, T.~A. 2010, \apj, 717, 196

\bibitem[\protect\citeauthoryear{{Lagache} et~al.}{{Lagache}
  et~al.}{2004}]{lagache2004}
{Lagache}, G., et~al. 2004, \apjs, 154, 112

\bibitem[\protect\citeauthoryear{{Lilly} et~al.}{{Lilly}
  et~al.}{1999}]{lilly1999}
{Lilly}, S.~J., {Eales}, S.~A., {Gear}, W.~K.~P., {Hammer}, F., {Le F{\`e}vre},
  O., {Crampton}, D., {Bond}, J.~R.,  \& {Dunne}, L. 1999, \apj, 518, 641

\bibitem[\protect\citeauthoryear{{Lonsdale} et~al.}{{Lonsdale}
  et~al.}{2004}]{lonsdale2004}
{Lonsdale}, C., et~al. 2004, \apjs, 154, 54

\bibitem[\protect\citeauthoryear{{Lucas} et~al.}{{Lucas}
  et~al.}{2003}]{lucas2003}
{Lucas}, R.~A., et~al. 2003, \aj, 125, 398

\bibitem[\protect\citeauthoryear{{Marchesini} et~al.}{{Marchesini}
  et~al.}{2009}]{marchesini2009}
{Marchesini}, D., {van Dokkum}, P.~G., {F{\"o}rster Schreiber}, N.~M., {Franx},
  M., {Labb{\'e}}, I.,  \& {Wuyts}, S. 2009, \apj, 701, 1765

\bibitem[\protect\citeauthoryear{{Marleau} et~al.}{{Marleau}
  et~al.}{2004}]{marleau2004}
{Marleau}, F.~R., et~al. 2004, \apjs, 154, 66

\bibitem[\protect\citeauthoryear{{Middelberg} et~al.}{{Middelberg}
  et~al.}{2008}]{middelberg2008}
{Middelberg}, E., {Norris}, R.~P., {Tingay}, S., {Mao}, M.~Y., {Phillips},
  C.~J.,  \& {Hotan}, A.~W. 2008, \aap, 491, 435

\bibitem[\protect\citeauthoryear{{Murphy}}{{Murphy}}{2009}]{murphy2009}
{Murphy}, E.~J. 2009, \apj, 706, 482

\bibitem[\protect\citeauthoryear{{Muzzin} et~al.}{{Muzzin}
  et~al.}{2010}]{muzzin2010}
{Muzzin}, A., {van Dokkum}, P., {Kriek}, M., {Labbe}, I., {Cury}, I.,
  {Marchesini}, D.,  \& {Franx}, M. 2010, ArXiv e-prints

\bibitem[\protect\citeauthoryear{{Norris} et~al.}{{Norris}
  et~al.}{2006}]{norris2006}
{Norris}, R.~P., et~al. 2006, \aj, 132, 2409

\bibitem[\protect\citeauthoryear{{Norris} et~al.}{{Norris}
  et~al.}{2007}]{norris2007}
{Norris}, R.~P., {Tingay}, S., {Phillips}, C., {Middelberg}, E., {Deller}, A.,
  \& {Appleton}, P.~N. 2007, \mnras, 378, 1434

\bibitem[\protect\citeauthoryear{{Palunas} et~al.}{{Palunas}
  et~al.}{2000}]{palunas2000}
{Palunas}, P., et~al. 2000, \apj, 541, 61

\bibitem[\protect\citeauthoryear{{Papovich} et~al.}{{Papovich}
  et~al.}{2004}]{papovich2004}
{Papovich}, C., et~al. 2004, \apjs, 154, 70

\bibitem[\protect\citeauthoryear{{Quadri} et~al.}{{Quadri}
  et~al.}{2007}]{quadri2007}
{Quadri}, R., et~al. 2007, \aj, 134, 1103

\bibitem[\protect\citeauthoryear{{Rieke} et~al.}{{Rieke}
  et~al.}{2004}]{rieke2004}
{Rieke}, G.~H., et~al. 2004, \apjs, 154, 25

\bibitem[\protect\citeauthoryear{{Sargent} et~al.}{{Sargent}
  et~al.}{2010}]{sargent2010}
{Sargent}, M.~T., et~al. 2010, \apjs, 186, 341

\bibitem[\protect\citeauthoryear{{Seymour} et~al.}{{Seymour}
  et~al.}{2007}]{seymour2007}
{Seymour}, N., et~al. 2007, \apjs, 171, 353

\bibitem[\protect\citeauthoryear{{Seymour} et~al.}{{Seymour}
  et~al.}{2010}]{seymour2010}
{Seymour}, N., {Symeonidis}, M., {Page}, M.~J., {Huynh}, M., {Dwelly}, T.,
  {McHardy}, I.~M.,  \& {Rieke}, G. 2010, \mnras, 402, 2666

\bibitem[\protect\citeauthoryear{{Shupe} et~al.}{{Shupe}
  et~al.}{2008}]{shupe2008}
{Shupe}, D.~L., et~al. 2008, \aj, 135, 1050

\bibitem[\protect\citeauthoryear{{Werner} et~al.}{{Werner}
  et~al.}{2004}]{werner2004}
{Werner}, M.~W., et~al. 2004, \apjs, 154, 1

\bibitem[\protect\citeauthoryear{{Williams} et~al.}{{Williams}
  et~al.}{2000}]{williams2000}
{Williams}, R.~E., et~al. 2000, \aj, 120, 2735

\bibitem[\protect\citeauthoryear{{Williams} et~al.}{{Williams}
  et~al.}{1996}]{williams1996}
{Williams}, R.~E., et~al. 1996, \aj, 112, 1335

\bibitem[\protect\citeauthoryear{{Yan} et~al.}{{Yan} et~al.}{2007}]{yan2007}
{Yan}, L., et~al. 2007, \apj, 658, 778

\bibitem[\protect\citeauthoryear{{Yun}, {Reddy}, \& {Condon}}{{Yun}
  et~al.}{2001}]{yun2001}
{Yun}, M.~S., {Reddy}, N.~A.,  \& {Condon}, J.~J. 2001, \apj, 554, 803

\end{thebibliography}

\begin{center}
\begin{deluxetable}{lcccrr}
\tablewidth{0pt}
\tabletypesize{\small}
\tablecaption{The MIPS 24$\mu$m catalog.}
\tablehead{\colhead{ID}  & \colhead{RA} & \colhead{Dec} & 
\colhead{$S_{24{\mu}m}$} & \colhead{$dS_{24{\mu}m}$} & \colhead{SNR}  }
\startdata
   MUSYC\_MIPSJ223100.6$-$603403 &  337.7524414 &  $-$60.5674782 &      882.0 &     46.2 &     31.8 \\
   MUSYC\_MIPSJ223105.6$-$603357 &  337.7733459 &  $-$60.5657158 &      355.8 &     15.0 &     15.2 \\
   MUSYC\_MIPSJ223106.3$-$603421 &  337.7760620 &  $-$60.5725441 &      719.7 &     15.1 &     32.2 \\
   MUSYC\_MIPSJ223110.7$-$603126 &  337.7944946 &  $-$60.5239792 &      717.2 &     55.2 &     28.5 \\
   MUSYC\_MIPSJ223111.0$-$603518 &  337.7959290 &  $-$60.5884361 &      205.6 &     11.3 &     10.2 \\

\enddata
\tablenotetext{.}{Table 1 is available in its entirety via the online version.}
\label{catalog}
\end{deluxetable}
\end{center}

\begin{center}
\begin{table*}[bt]
\hfill{}
\begin{tabular}{ccccc}
\hline 
Range in $S_{24{\mu}m}$ & $\langle S_{24{\mu}m}  \rangle$ & N & corrected N & $(dN/dS) S^{2.5}$ \\
    ($\mu$Jy)                          &               ($\mu$Jy)                            &     &                       & [mJy$^{1.5}$  deg$^{-2}$] \\ \hline 
110 -- 134  & 115 & 24 & 121.4 & 368.8 $\pm$ 75.3 \\
134 -- 163  & 145 & 65 & 149.4 & 650.7 $\pm$ 80.7 \\
163 -- 198  & 174 & 70 & 106.8 & 585.4 $\pm$ 70.0 \\
198 -- 241  & 213 & 83 & 97.5 &   738.0 $\pm$ 81.0 \\
241 -- 293  & 254 & 70 & 76.0 &   753.0 $\pm$ 90.0 \\
293 -- 356 & 316  & 54 & 58.0 &    788.1 $\pm$107.2 \\
356 -- 433 & 388 & 37 & 38.4 &     706.3 $\pm$ 116.1 \\
433 -- 526 & 454 & 29 & 29.7 &    659.0 $\pm$ 122.4 \\
526 -- 640 & 590 & 13 & 13.2 &    444.6 $\pm$ 123.3 \\
640 -- 779 & 746 & 14& 14.3 &     712.5 $\pm$ 190.4 \\ \hline
\end{tabular} 
\hfill{}
\caption{The MIPS 24$\mu$m source count for Hubble Deep Field South. $N$ is the raw number of galaxies in the bin, and corrected $N$ is the number corrected for incompleteness. The median flux density of the bin, $\langle S_{24{\mu}m}  \rangle$, has been corrected for flux boosting. }
\end{table*}
\end{center}

\begin{center}
\begin{table*}[bt]
\hfill{}
\begin{tabular}{ccrrcrrr}
\hline 
$z$ & median $z$ & $N_{det}$ & $N_{stack}$ & stacked $S_{1.4 GHz}$ ($\mu$Jy) & stacked  $q_{24}$ & average  $q_{24}$  \\  \hline
$0  < z < 0.25$ & 0.149 & 8 & 25 & 15.2 $\pm$ 4.1 & $1.38^{+0.10}_{-0.14}$ & $1.09^{+0.06}_{-0.07}$ \\
$0.25  < z < 0.50$ & 0.389 & 19 & 54 & 23.8 $\pm$ 3.5 & $1.18^{+0.06}_{-0.07}$ & $0.93^{+0.05}_{-0.05}$ \\
$0.50  < z < 0.75$ & 0.580 & 14 & 65 & 22.8 $\pm$ 5.2 & $1.05^{+0.09}_{-0.11}$ & $0.91^{+0.07}_{-0.08}$ \\
$0.75  < z < 1.00$ & 0.890 & 3 & 57 & 16.4 $\pm$ 5.0 & $1.14^{+0.12}_{-0.16}$ & $1.14^{+0.10}_{-0.14}$ \\
$1.00  < z < 1.50$ & 1.270 & 10 & 72 & 18.2 $\pm$ 3.0 & $1.11^{+0.07}_{-0.08}$ & $0.86^{+0.06}_{-0.06}$ \\
$1.50  < z < 2.50$ & 1.787 & 13 & 61 & 12.0 $\pm$ 2.9 & $1.27^{+0.10}_{-0.12}$ & $0.98^{+0.05}_{-0.06}$ \\
no redshift & \nodata & 16 & 73 & 18.7 $\pm$ 3.6 & $1.12^{+0.08}_{-0.10}$ & $0.89^{+0.06}_{-0.06}$ \\ \hline
\end{tabular} 
\hfill{}
\caption{Summary of radio stacking results of MIPS sources. $N_{det}$ is the number of MIPS sources in each bin which is detected in the radio image. $N_{stack}$ is the number of MIPS sources not detected in the radio image and therefore the number stacked in this analysis, resulting in the stacked $S_{1.4 GHz}$ listed in column 5. Stacked $q_{24}$ is the IR-radio flux density ratio for the stacked sample. Average $q_{24}$ is the IR-radio flux density ratio for all MIPS sources in the redshift bin, obtained by combining the detections with the stacked results. }
\end{table*}
\end{center}

\begin{center}
\begin{table*}[bt]
\footnotesize
\begin{minipage}[b]{1\linewidth}
\begin{tabular}{ccrrccrc}
\hline 
radio ID & $S_{24{\mu}m}$ limit & $S_{1.4 GHz}$ & $S_{3.6{\mu}m}$ & $q24$& $S_{1.4GHz}$/$S_{3.6{\mu}m}$& $z_{phot}$ & $\alpha^{2.5 GHz}_{1.4GHz}$ \footnotemark[1]\\ 
    & (5$\sigma$, $\mu$Jy) & (mJy) & ($\mu$Jy) & \\ \hline
ATHDFS\_J223224.5$-$604113 & 130 & 4.108 & 112 & $<-1.50$ & 37 & 0.746 & $-0.48 \pm 0.02$ \\ 
ATHDFS\_J223245.6$-$603857 & 100 & 0.843 & 31.2 & $<-0.93$ & 27 & 1.548 & $-0.69 \pm 0.06$ \\
 ATHDFS\_J223205.9$-$603857 & 70 &   0.254 & $<2$ & $<-0.56$ & $>$127 & \nodata\footnotemark[2] & $-0.80 \pm 0.29$\\
 ATHDFS\_J223350.0$-$603741& 80 & 0.670 & 156 & $<-0.93$ & 4.4 & 0.645 & $-0.36 \pm 0.05$ \\
ATHDFS\_J223232.8$-$603737 & 65 & 0.645 & 92.6 & $<-1.00$ & 7.0 & 1.13 & $-0.51 \pm 0.08$ \\
ATHDFS\_J223350.5$-$603503 & 110 & 1.252 & $>$4\footnotemark[3] & $<-1.06$ & $<$310\footnotemark[3] & 0.417 & $-0.85 \pm 0.05$\\
ATHDFS\_J223258.5$-$603346 & 95 & 1.010 & 21.31 & $<-1.03$ & 47.4 & \nodata & $-0.65 \pm 0.05$ \\
ATHDFS\_J223308.6$-$603251 & 75 & 0.821 & 101 & $<-1.04$ & 8.1 & 0.746 & $-0.86 \pm 0.03$ \\
ATHDFS\_J223113.5$-$603147 & 115 & 0.831 & \nodata\footnotemark[4] & $<-0.86$ & \nodata\footnotemark[4] & \nodata\footnotemark[2] & \nodata\\ \hline
\end{tabular} 
\caption{Summary of radio sources that have no MIPS 24 $\mu$m counterpart and hence are candidate infrared-faint radio sources. Only ATHDFS\_J223205.9$-$603857 is faint enough at 3.6$\mu$m to meet the original IFRS definition of Norris et al. 2006. All of these sources have $q_{24}$ values suggestive of radio-loud AGN. The 3.5 $\mu$m data is from \cite{marchesini2009}.}
\footnotetext[1]{Radio spectral indices are from \cite{huynh2007}, and $S \propto \nu^\alpha$}
\footnotetext[2]{Not detected in MUSYC optical imaging.}
\footnotetext[3]{IRAC image of source has artefacts, so photometry is inaccurate.}
\footnotetext[4]{Out of IRAC coverage.}
\end{minipage}
\end{table*}
\end{center}

\begin{figure}[htb]
\includegraphics[width=9.5cm]{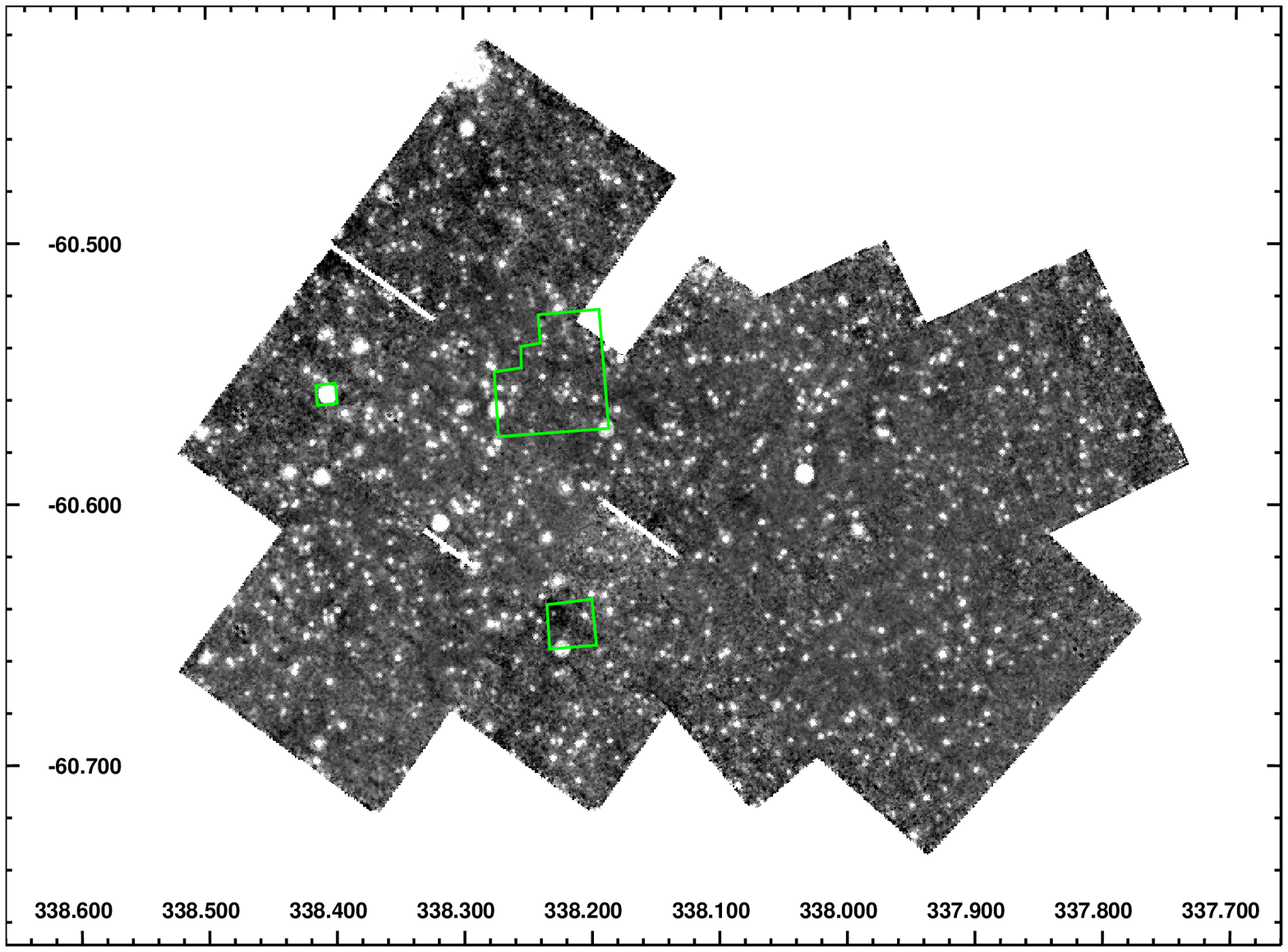}
\includegraphics[width=7.2cm]{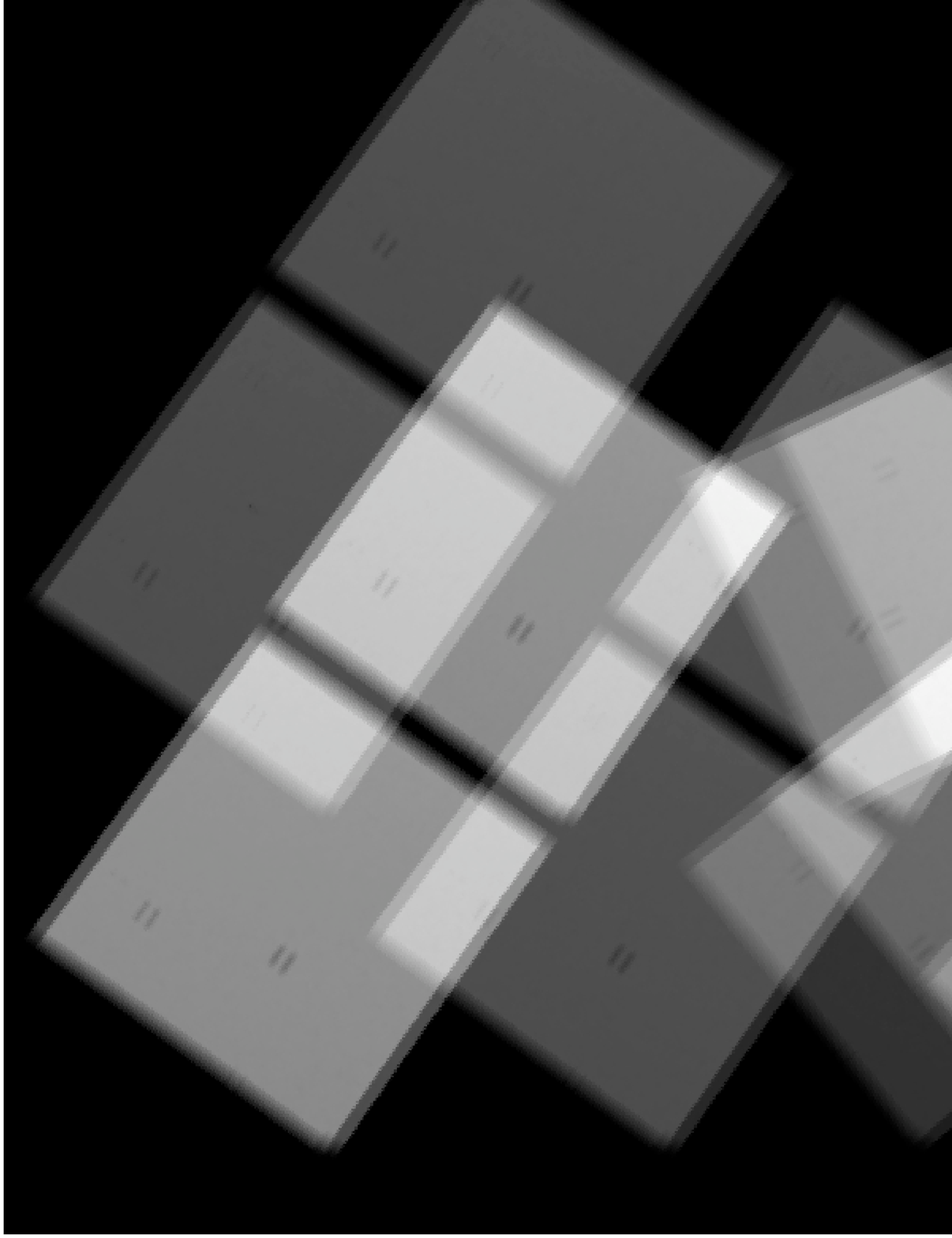}
\caption{LEFT: The MIPS 24 micron image of the Hubble Deep Field South region. The green chevron figure marks the area covered by the original HST WFPC2 deep field, the squares to the east and south are the HST STIS and NICMOS deep fields, respectively. RIGHT: The coverage map of the MIPS 24 micron image, white indicates most coverage and black indicates no coverage.}
\end{figure}

\begin{figure*}[tb]
\includegraphics[width=7.8cm]{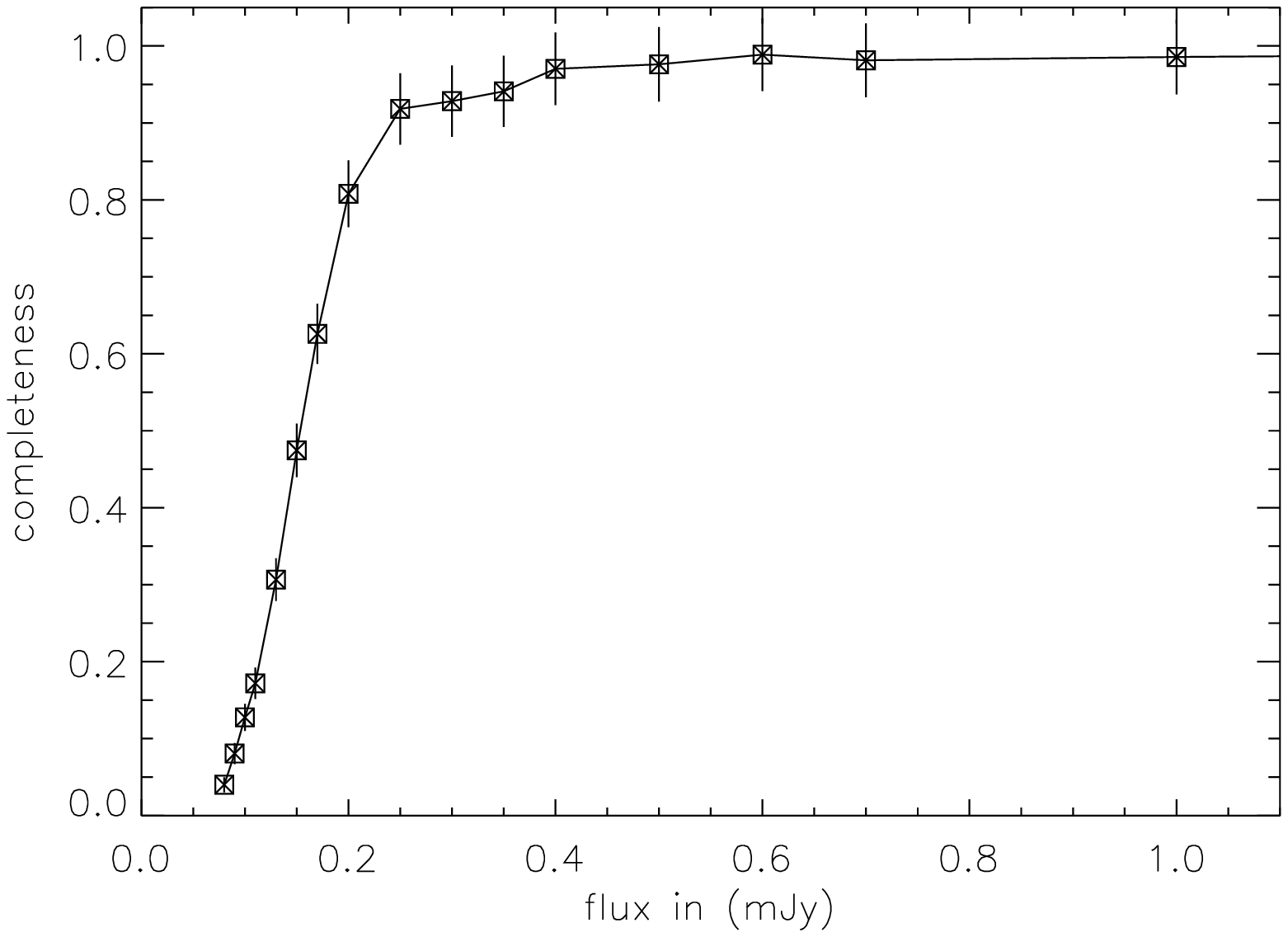}
\includegraphics[width=7.8cm]{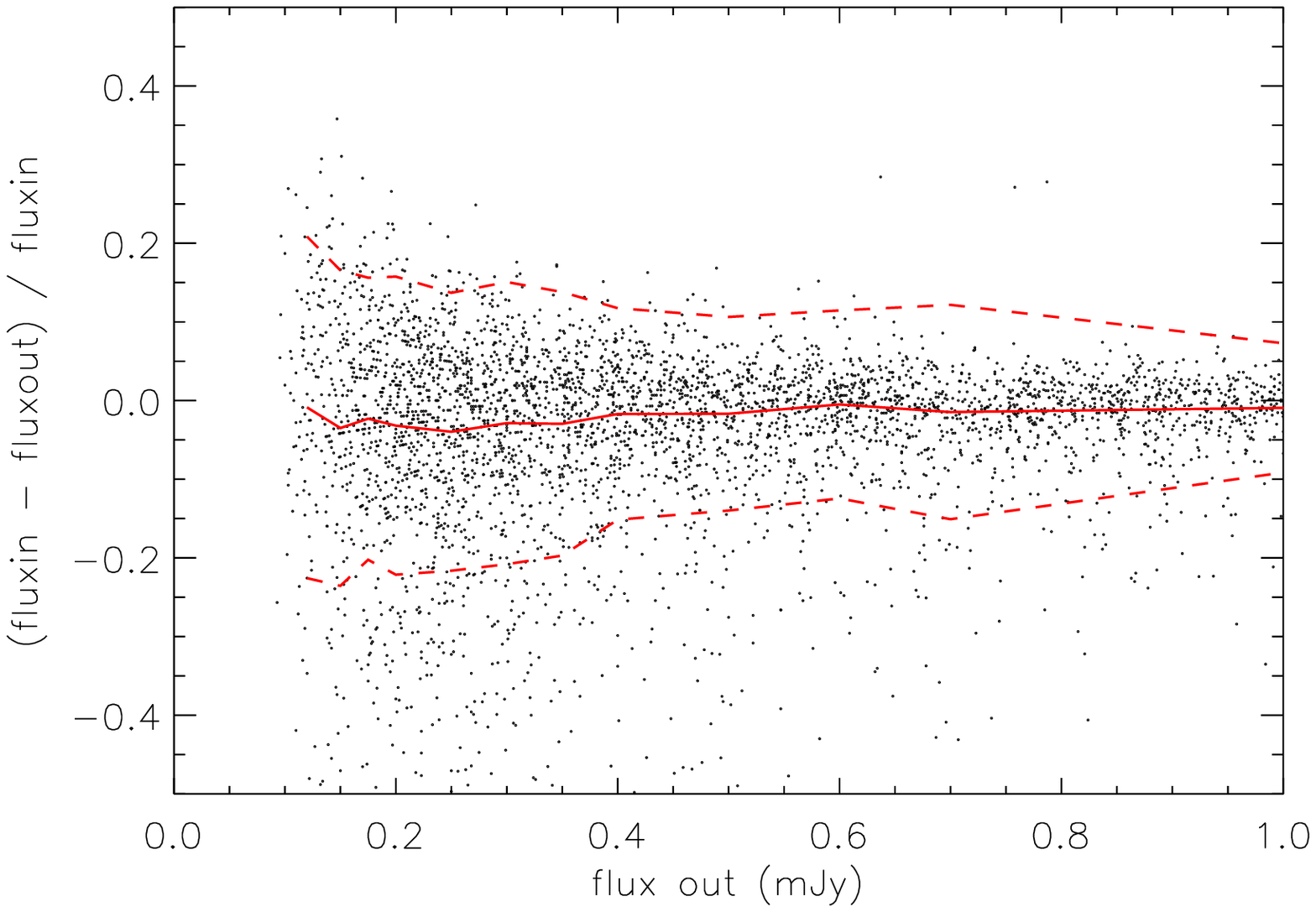}
\caption{LEFT: Completeness as a function of input flux density, as derived from the Monte-Carlo simulations. Completeness is the number of extracted sources divided by number of input sources. RIGHT: The distribution of (input flux density - output flux density)/input flux density as a function of output flux density for the simulated sources. The solid red line is the mean of the simulation and the dashed lines mark the 1 sigma upper and lower bounds.}
\end{figure*}

\begin{figure*}[htb]
\includegraphics[width=7.5cm]{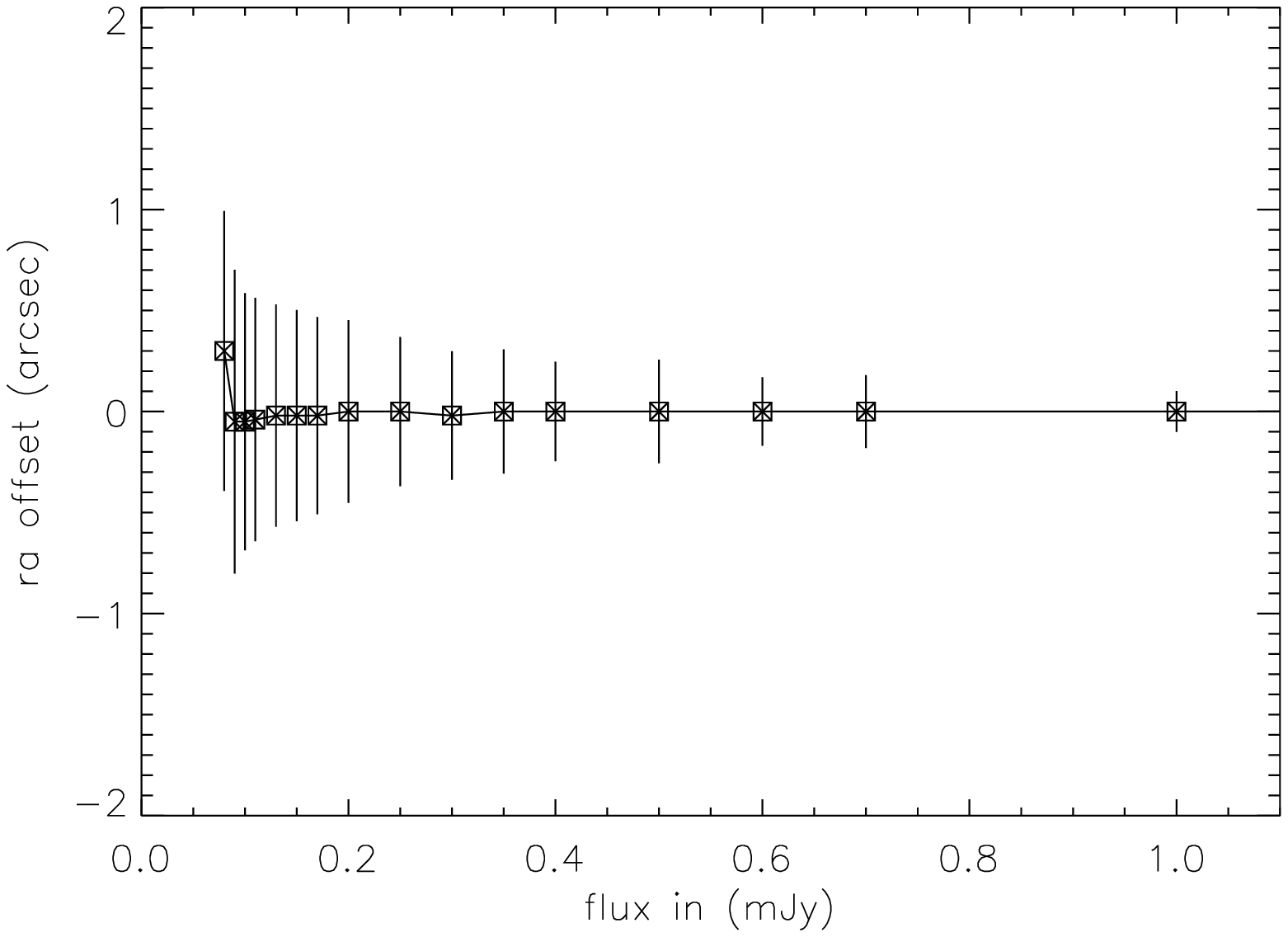}
\includegraphics[width=7.5cm]{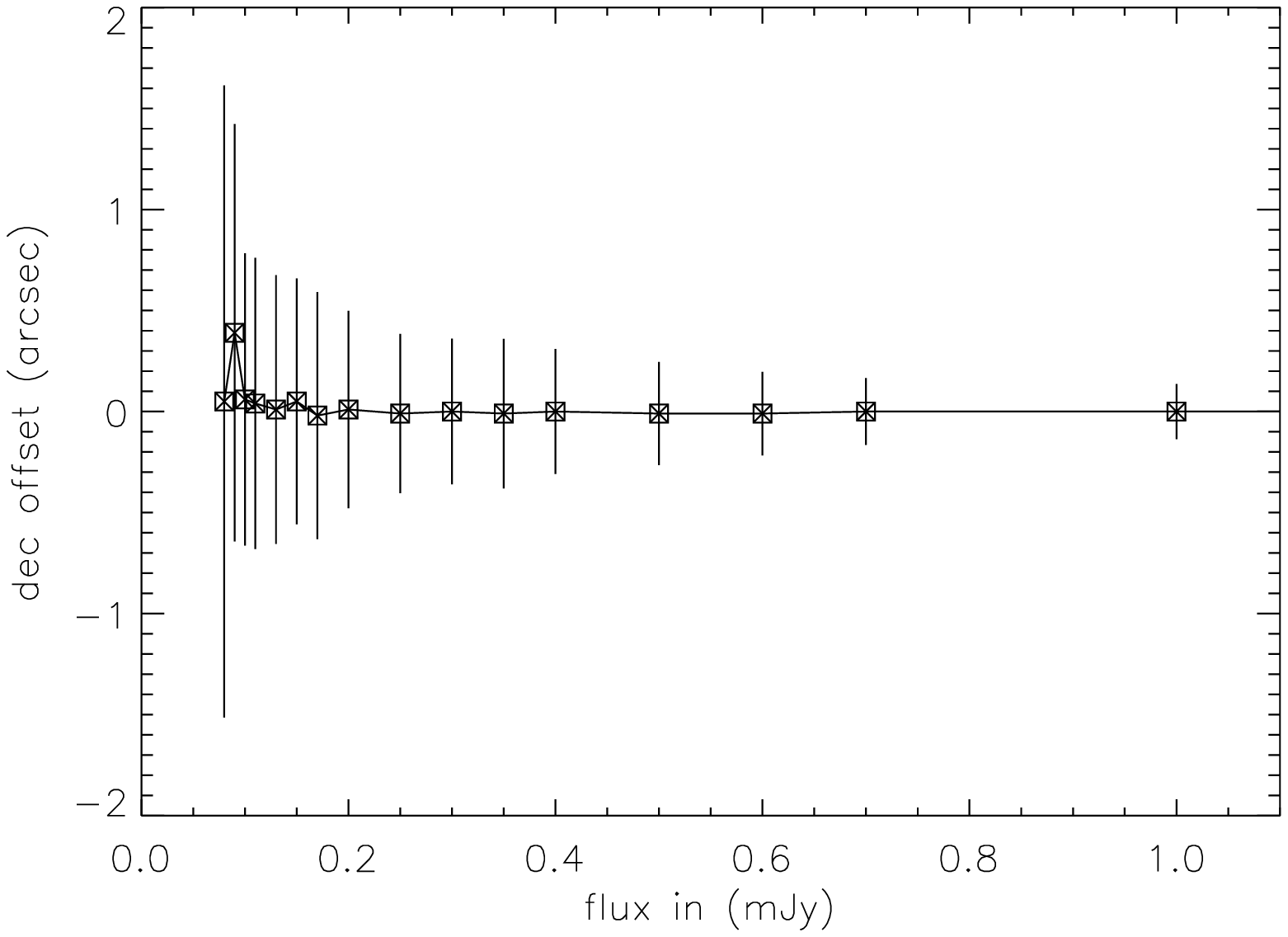}
\caption{The offset in RA (left) and Dec (right) between the recovered positions of sources in the simulation and the true input positions, as a function of input flux density. The error bars mark the 1 sigma uncertainty in the position as a function of input flux density. }
\end{figure*}

\begin{figure}[bt]
\includegraphics[width=12cm]{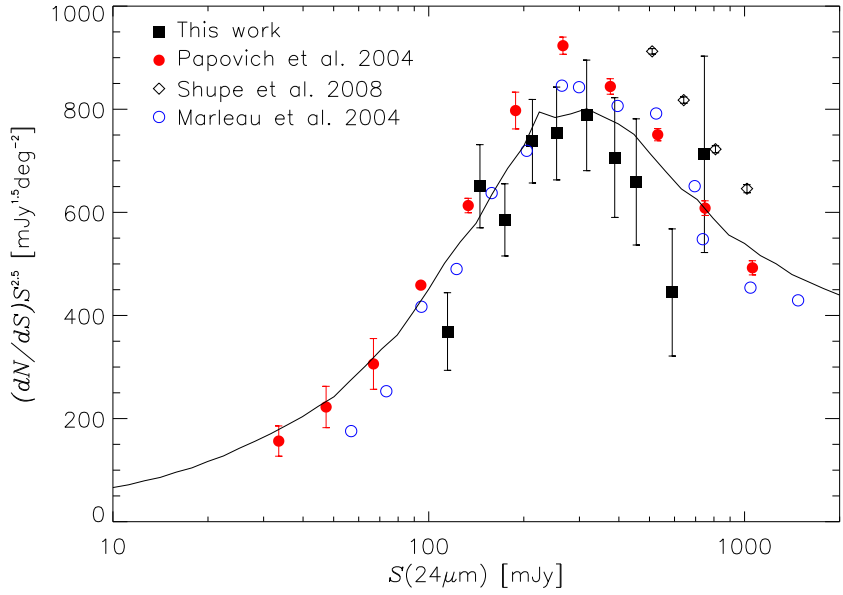}
\caption{The Euclidean normalized differential 24 $\mu$m source counts in the Hubble Deep Field South (black squares). These counts have been corrected for incompleteness and flux boosting as described in Section 4.2. The solid line shows the prediction from the model by Lagache et al. 2004. Counts from other deep surveys in the literature are plotted for comparison. }
\end{figure}

\begin{figure}[bt]
\includegraphics[width=7.5cm]{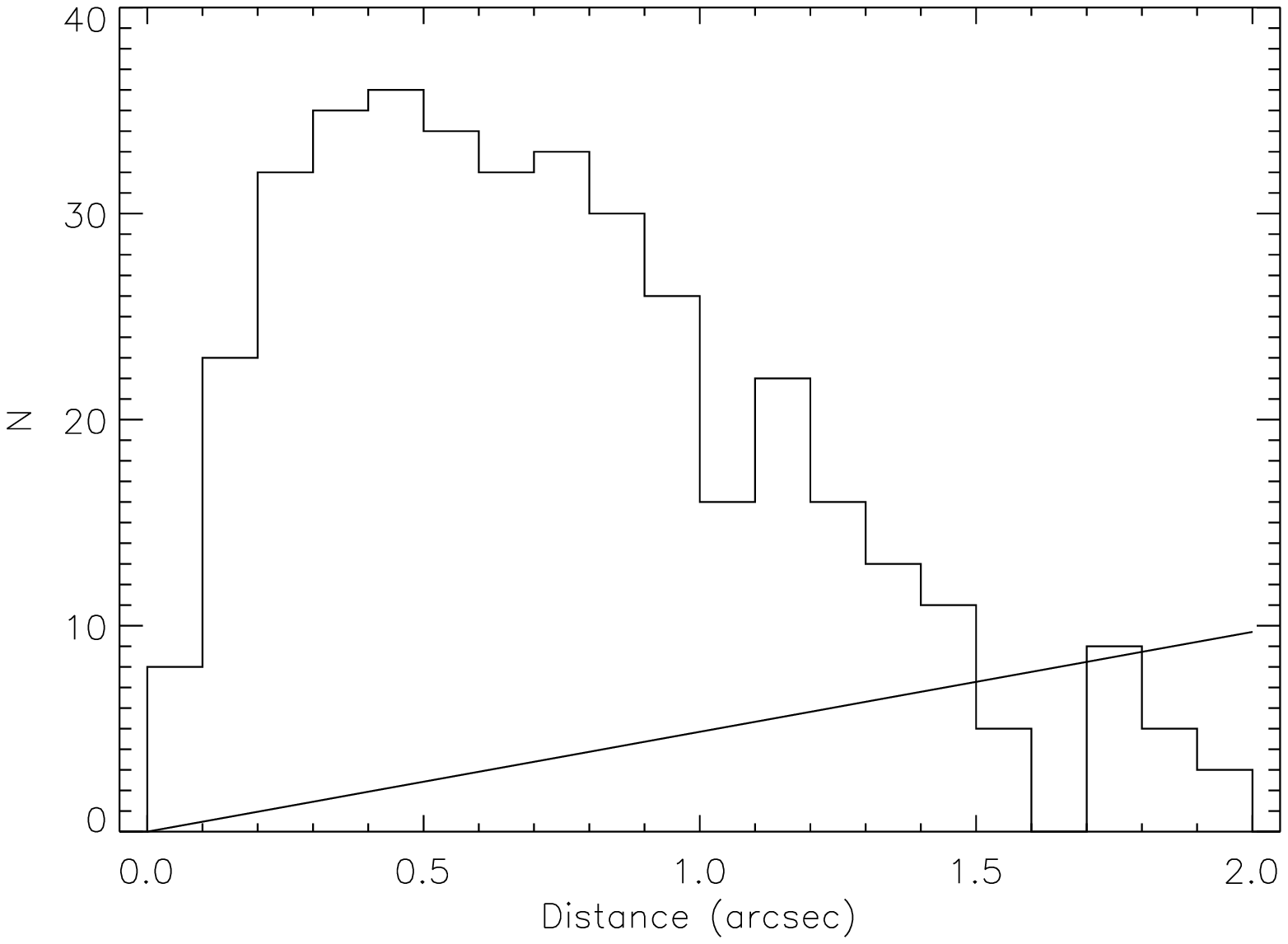}
\includegraphics[width=7.5cm]{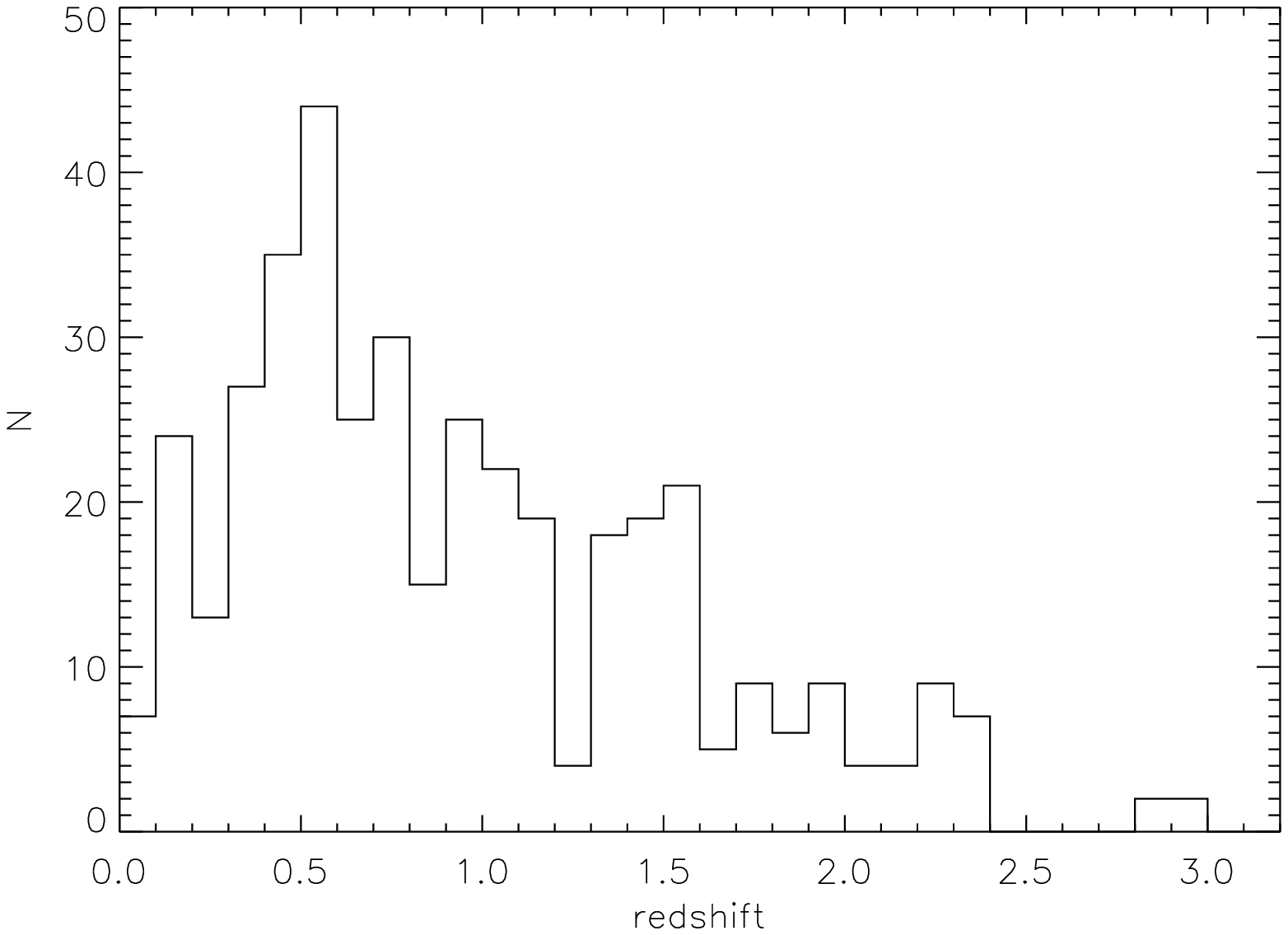}
\caption{LEFT: Histogram of the total offsets between the MUSYC optical candidate and the 24 $\mu$m source. Most MIR sources have an optical counterpart within 1 arcsec. The solid line is the expected number of coincident sources with the same offset, assuming the MUSYC source density of 56/arcmin$^2$. RIGHT: The redshift distribution of the 406 MIPS sources in HDFS with a MUSYC optical counterpart. There are spectroscopic redshifts for 22 sources.}
\end{figure}

\begin{figure}[bt]
\includegraphics[width=14cm]{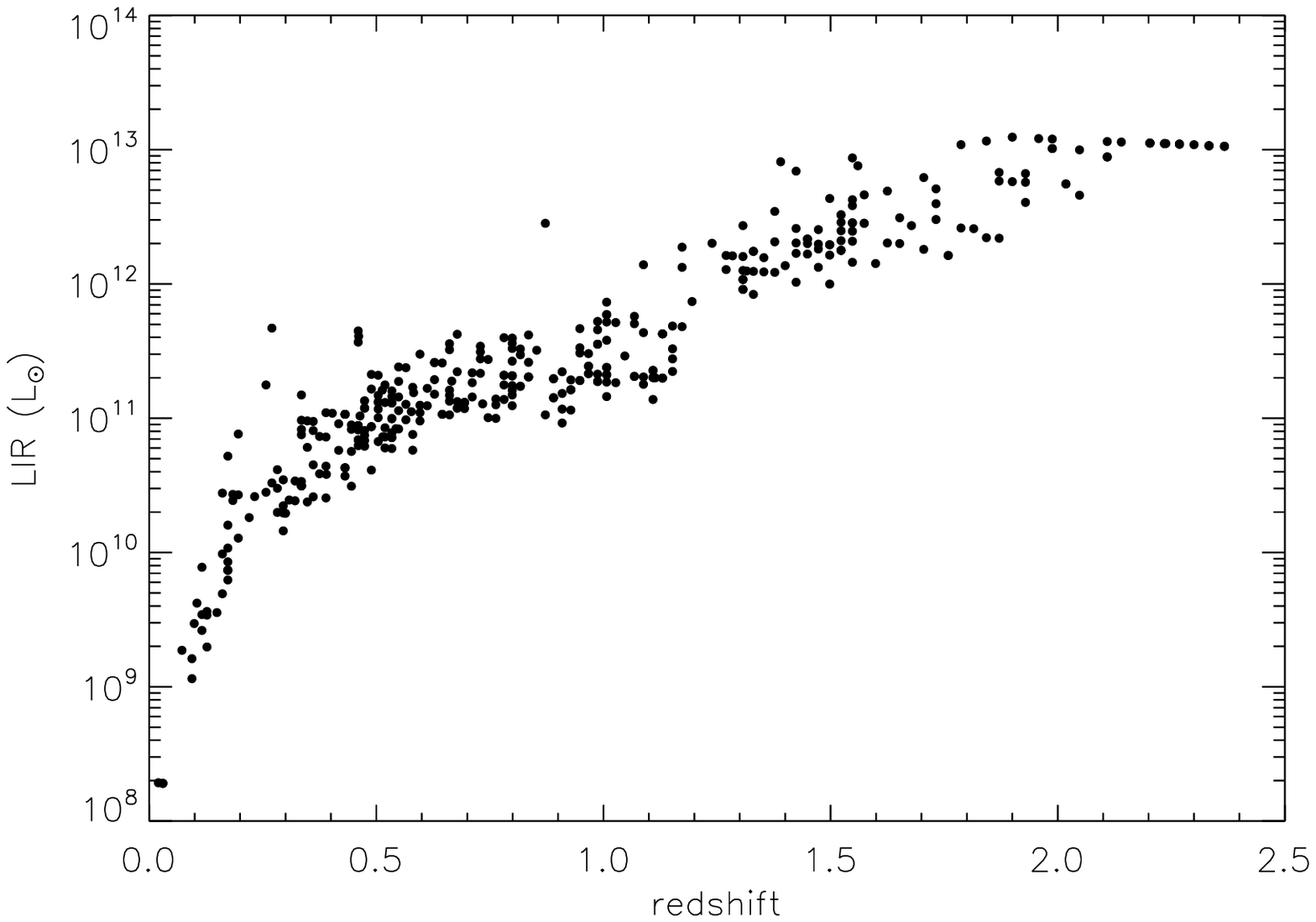}

\vspace{-6.2cm}\hspace{6.5cm}
\includegraphics[width=6.6cm]{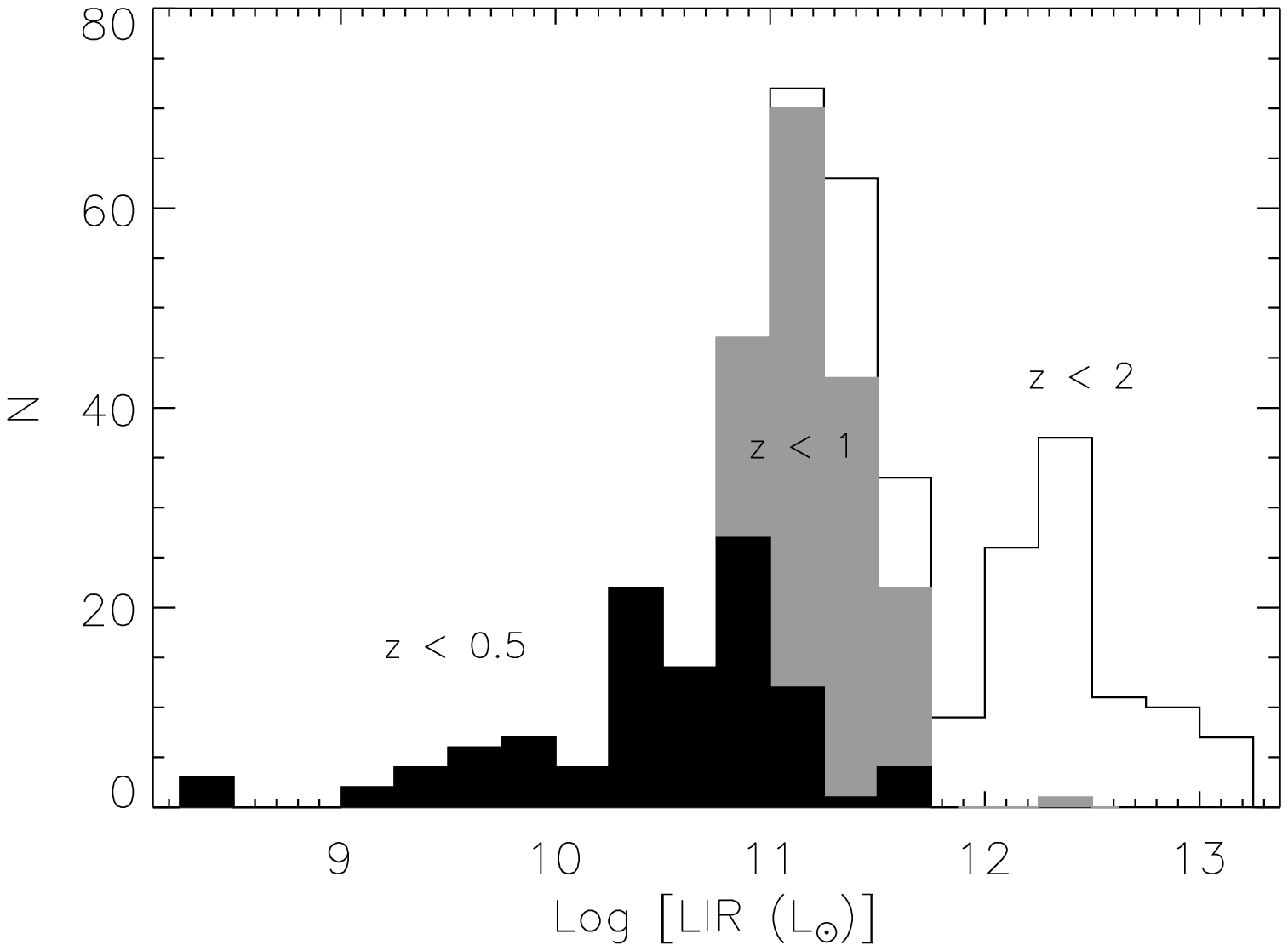}
\vspace{0.7cm}
\caption{Total IR luminosities of the 358 MIPS sources with redshift information. They were derived by fitting to the luminosity dependent SEDs of Chary and Elbaz (2001). {\em Inset} The corresponding IR luminosity histograms of the sample for $z < 0.5$ (black), $z < 1$ (grey) and $z < 2$ (white). }
\end{figure}

\begin{figure*}[bt]
\includegraphics[width=15cm]{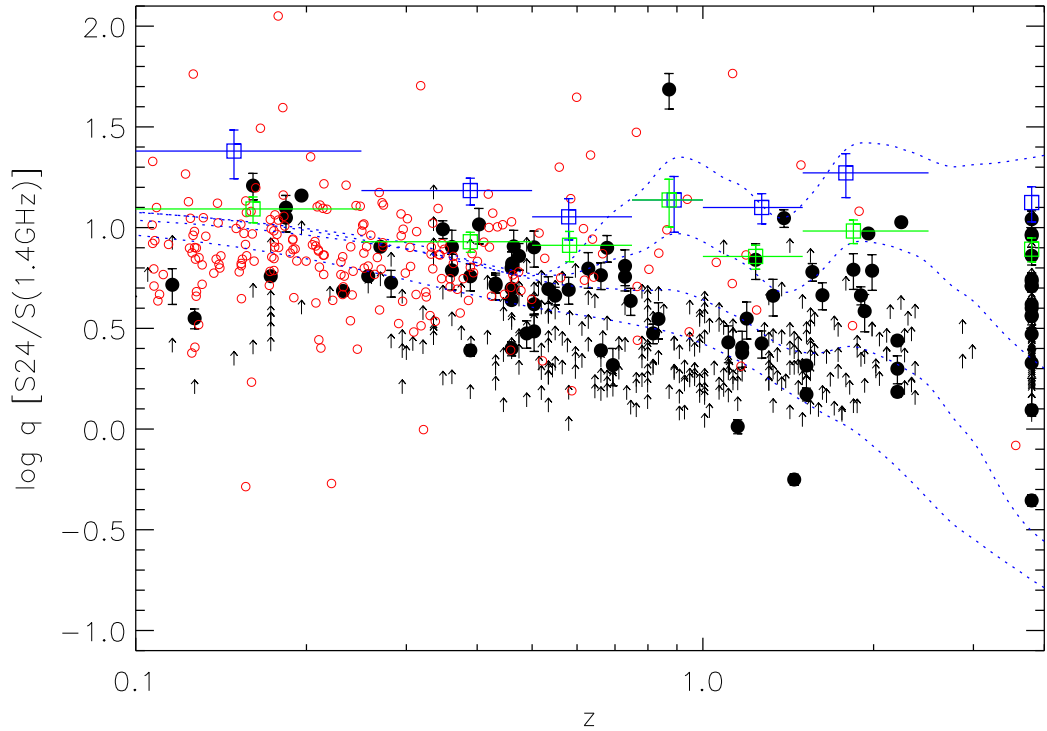}
\caption{The observed 24 $\mu$m versus radio flux density ratio ($q_{24}$) as a function of redshift. HDFS sources detected in MIR and radio are marked as black circles. MIR only sources not detected in the radio are plotted as lower limits assuming the conservative radio detection limit of 100 $\mu$Jy. Sources with no redshift information are plotted at $z = 3.8$. 
Red circles are results from the xFLS (Frayer et al. 2006). The blue squares are average $q_{24}$ for the MIR sources undetected in the radio, where the average radio flux density is derived from a stacking analysis. The green squares are the average $q_{24}$ for all MIR sources, i.e. combining the blue squares with radio detections. The dotted blue lines are SED tracks from Chary and Elbaz (2001) for galaxies with total infrared luminosities of $10^{9}$, $10^{11}$, $10^{12}$ and $10^{13}$ $L_{\odot}$, going from top to bottom.}
\end{figure*}

\begin{figure*}[bt]
\includegraphics[width=2.2cm]{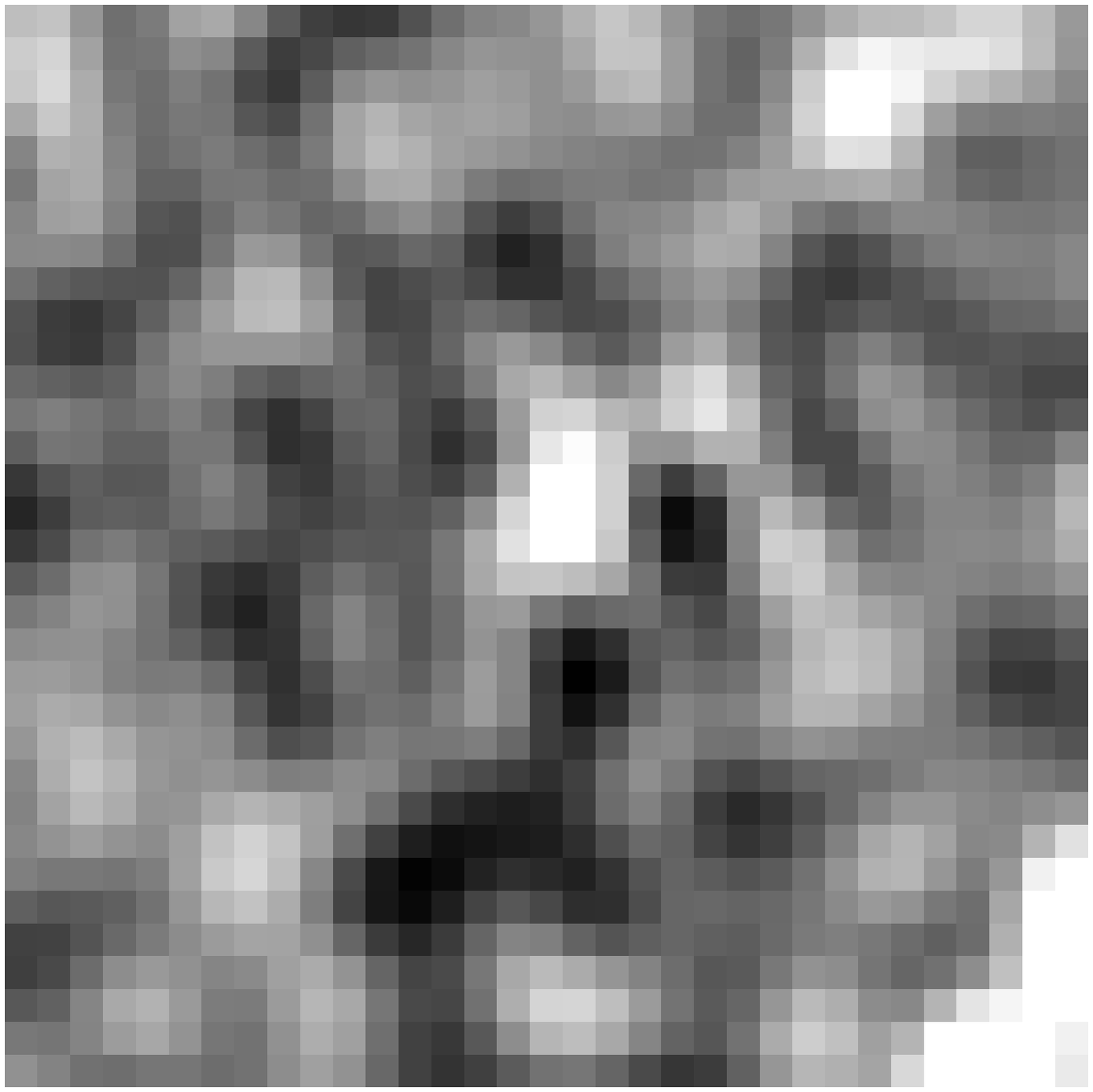}
\includegraphics[width=2.2cm]{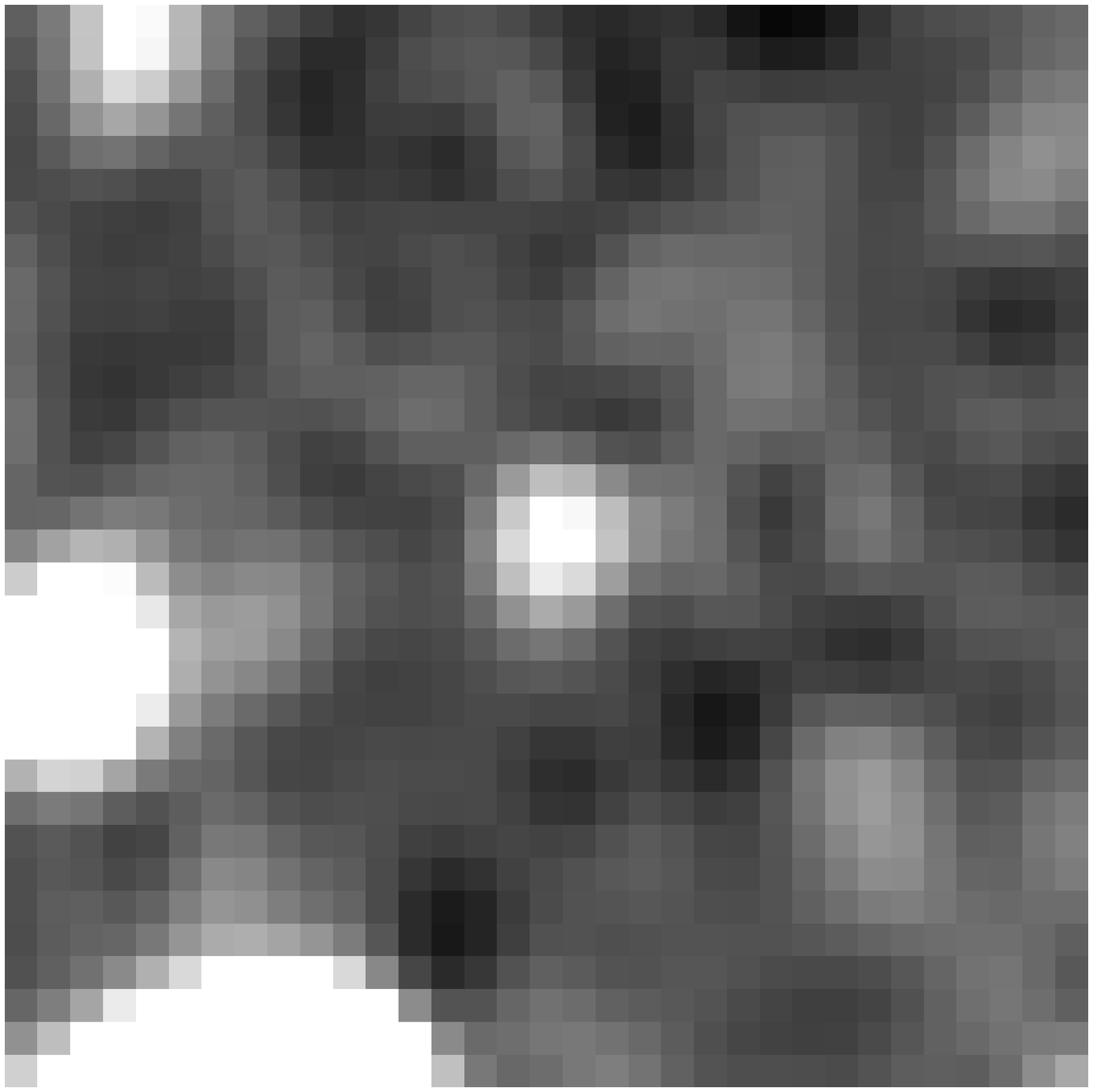}
\includegraphics[width=2.2cm]{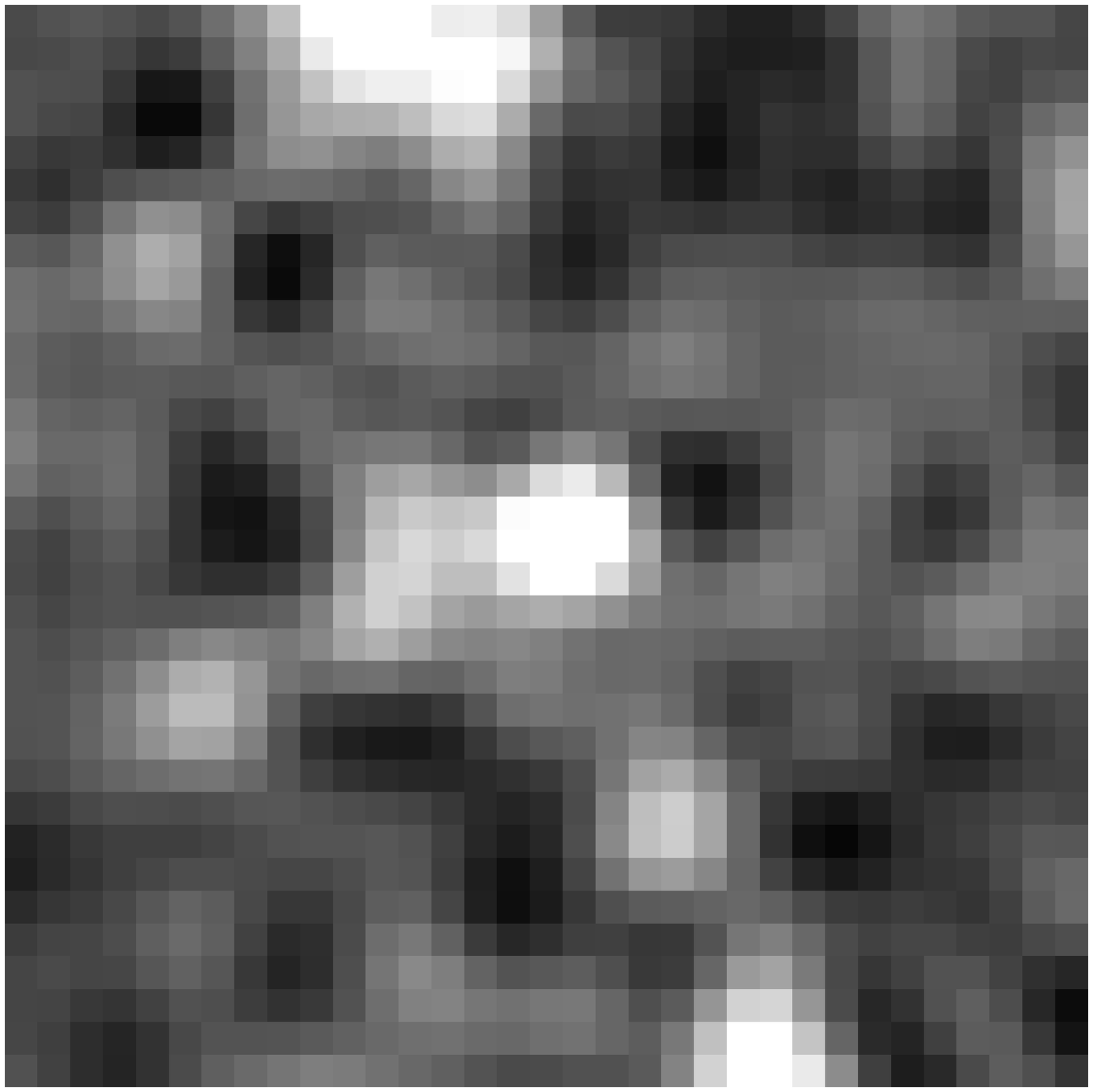}
\includegraphics[width=2.2cm]{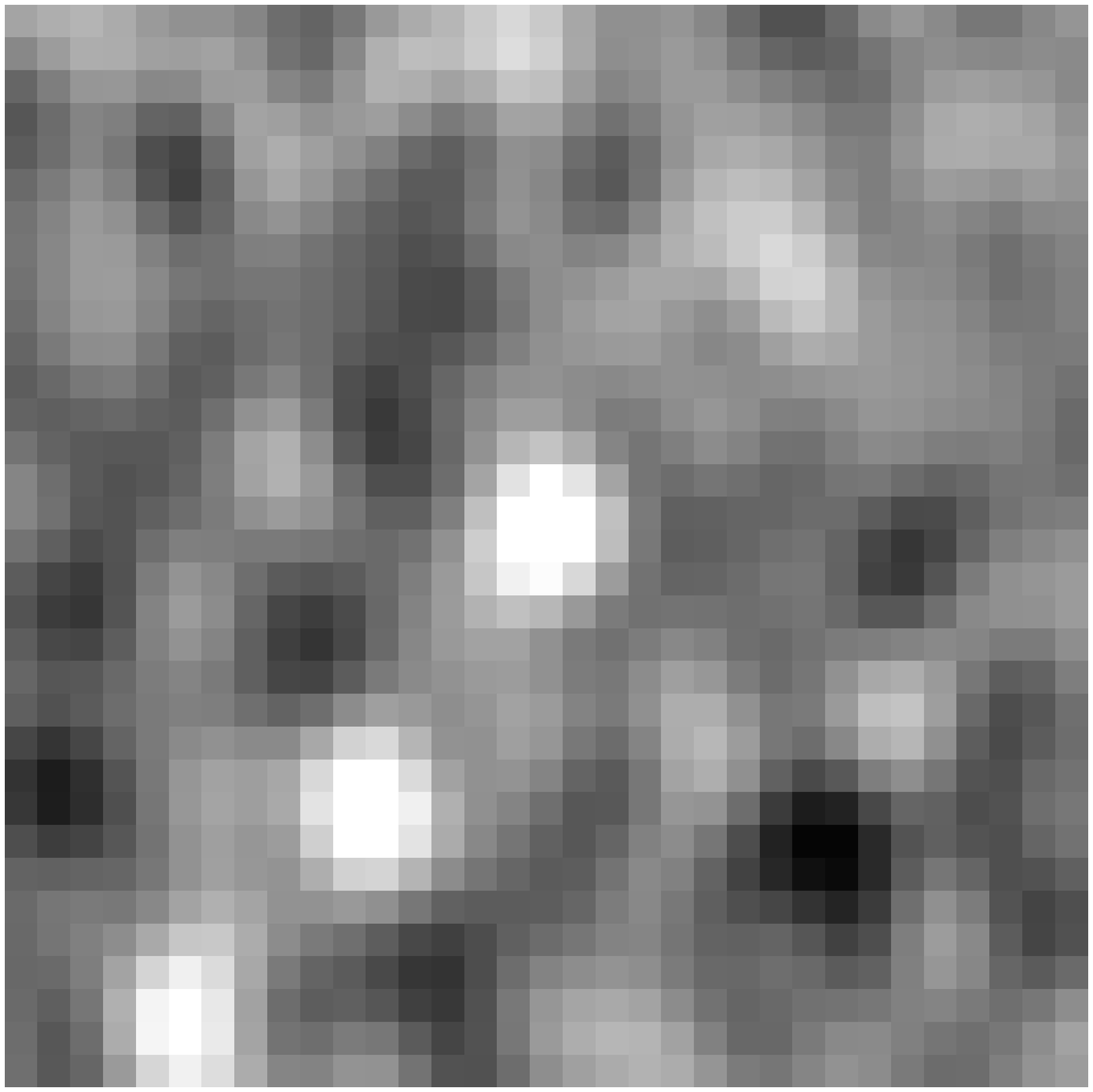}
\includegraphics[width=2.2cm]{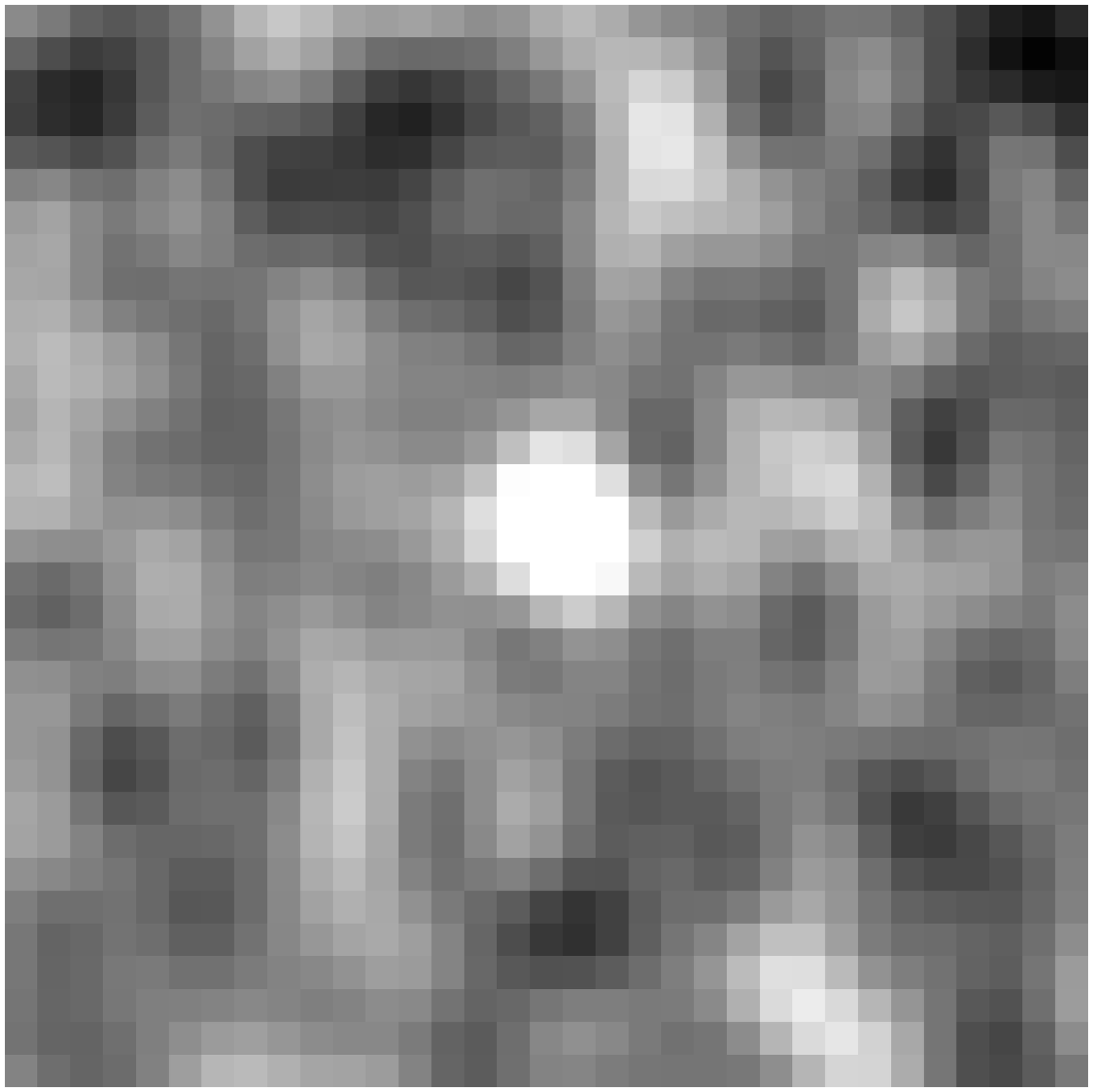}
\includegraphics[width=2.2cm]{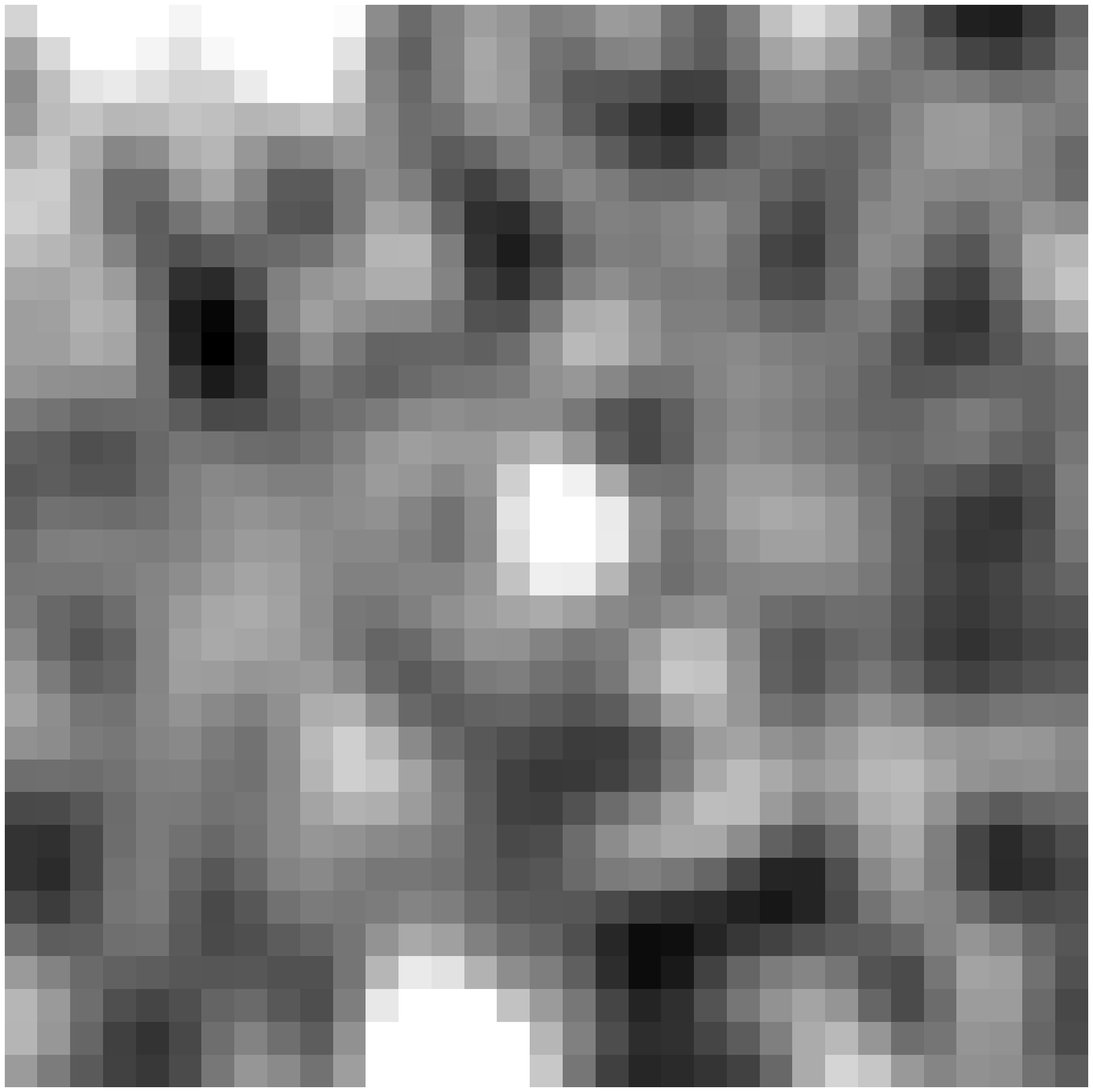}
\includegraphics[width=2.2cm]{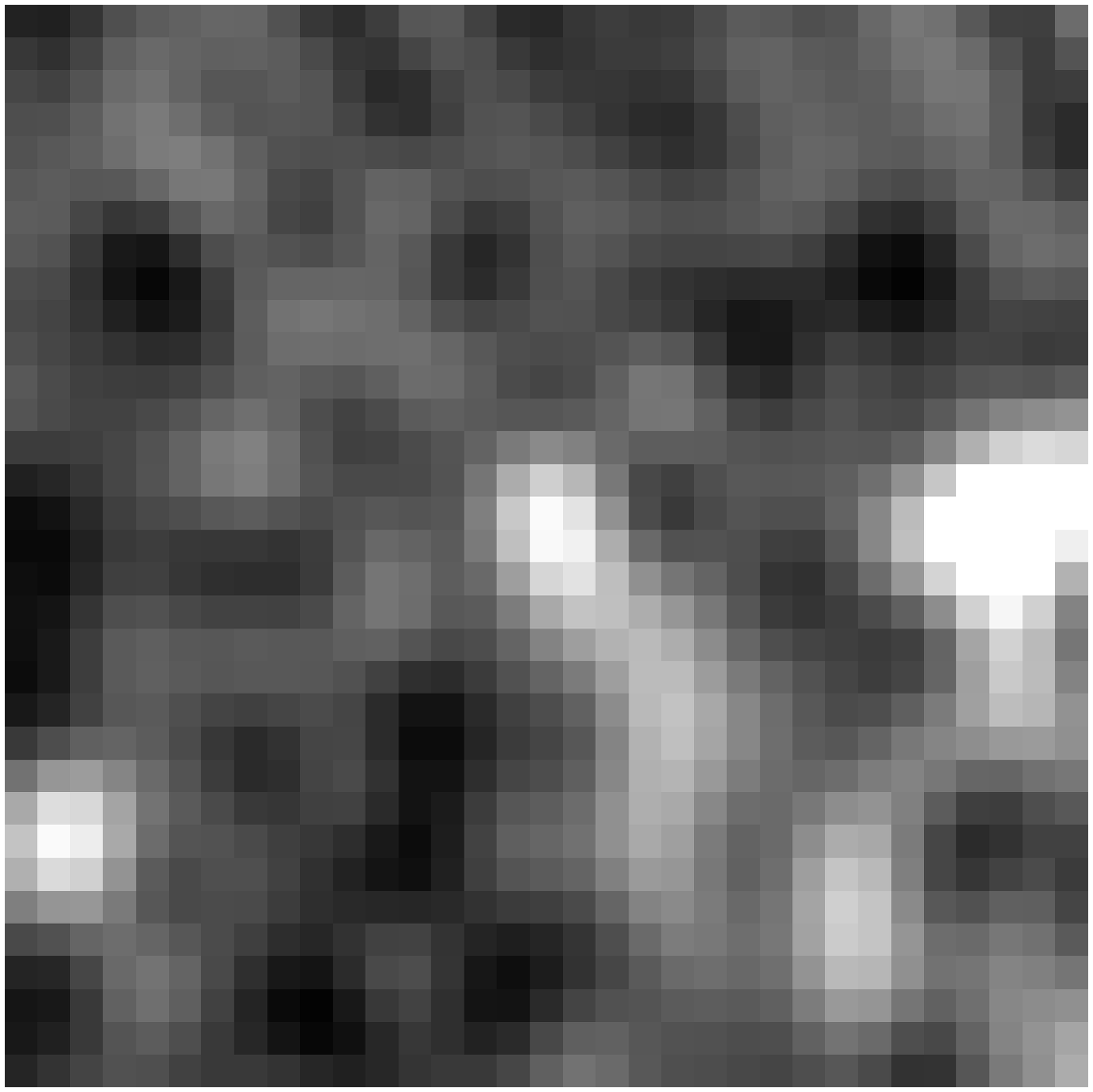}
\vspace*{-2mm}
{\tiny 
\begin{tabbing}
\hspace*{0.3cm} \= \hspace*{2.2cm} \= \hspace*{2.2cm} \= \hspace*{2.2cm}   \= \hspace*{2.2cm}  \= \hspace*{2.2cm}  \= \hspace*{2.9cm}  \= \hspace*{2.9cm}  
 \kill\> 
$0 < z < 0.25$ \> $0.25 < z < 0.50$  \>   $0.50 < z < 0.75$ \> $0.75 < z < 1.00$ \> $1.00 < z < 1.50$ \> $1.50 < z < 2.50$ \>  no $z$\\
\end{tabbing}
}
\caption{Postage stamps of the radio stacks of MIPS sources, for the redshift bins as shown. The stacks are  1 $\times$ 1 arcmin in size. In all cases there is a significant radio flux at the center of the stack, which is reported in Table 3.}
\end{figure*}

\end{document}